\documentclass{article}
\usepackage{iclr2027_conference,times}
\iclrfinalcopy
\usepackage{amsmath,amssymb,graphicx,booktabs}
\usepackage{xcolor}
\usepackage{algorithm,algpseudocode}
\usepackage{wrapfig}
\usepackage{placeins}
\usepackage{hyperref}
\usepackage{url}
\usepackage{enumitem}
\usepackage{pifont}
\makeatletter
\let\DSHOriginalMaketitle\@maketitle
\def\@maketitle{{\def\lhead##1{\fancyhead[L]{Preprint}}\DSHOriginalMaketitle}}
\makeatother

\newcommand{\revised}[1]{#1}
\title{Coarse-to-Fine MOF Generation with Periodic U-DiT and Verifier-Guided Linker Editing}
\author{Yuhan Li$^{1}$ \quad Zhaoying Zhang$^{1}$ \quad Xu Wang$^{1,2}$ \quad Binwu Wang$^{1,2}$ \quad Yang Wang$^{1,2,3,*}$\\[4pt]
\mdseries\normalsize
$^{1}$University of Science and Technology of China (USTC), Hefei, China\\
$^{2}$Suzhou Institute for Advanced Research, USTC, Suzhou, China\\
$^{3}$State Key Laboratory of Precision and Intelligent Chemistry, USTC, Hefei, China\\[2pt]
$^{*}$Corresponding author: \texttt{angyan@ustc.edu.cn}}
\begin{document}
\maketitle
\begin{abstract}
Controlling the pore-limiting diameter (PLD) of metal--organic frameworks (MOFs) requires connecting continuous pore geometry to discrete chemical assembly. We introduce CoFi-MOF, a coarse-to-fine framework that combines PLD-conditioned periodic diffusion, learned topology-aware structural realization, geometric verification and local linker editing. A Periodic U-DiT generates pore fields in a signed-distance-field latent space. A learned Constructor proposes building-block assignments, which are assembled and ranked by measured PLD. Multi-Edit-Former then proposes topology-preserving linker modifications, with reassembly and measurement determining the final structure. Under a fixed-upstream protocol with 25 measured realizations per attempt, CoFi-MOF achieves 86.20\% Hit@0.5 across PLD targets of 3--7~\AA, compared with at most 76.00\% for the evaluated neural pipelines. CoFi-MOF concentrates chemically distinct candidates around each requested aperture, supplying measured shortlists for downstream physical validation and experimental selection. We also expand PORMAKE's carboxylate-based linker library and introduce MOF-Edit-DB, a resource of 138,771 framework records with 108,145 matched parent--edit comparisons. Together, these contributions support target-directed exploration of small-aperture MOFs and learning from measured structural edits.

\end{abstract}
\section{Introduction}

Metal--organic frameworks (MOFs) have emerged at the frontier of porous-material design because their modular construction offers extensive control over both chemical composition and periodic geometry~\citep{yaghi2003reticular,furukawa2013chemistry}. This tunability has made MOFs promising candidates for gas storage, molecular separation, and catalysis~\citep{furukawa2013chemistry}. However, the enormous number of possible combinations among nodes, linkers, and network topologies makes exhaustive exploration through experimental synthesis or atomistic simulation impractical. This challenge has catalyzed the development of generative models that learn directly from existing MOF structures and propose candidates at substantially greater scale~\citep{yao2021smvae}. Recent advances include coarse-grained diffusion for generating novel and geometrically plausible frameworks in MOFDiff~\citep{fu2024mofdiff}, torsion-aware flow matching for flexible linkers and novel building blocks in MOFFlow-2~\citep{kim2025mofflow2}, pore-property-conditioned field generation and assembly in MOFFUSION~\citep{park2025moffusion}, and physical-reward optimization for targeted design in PRO-MOF~\citep{liu2026promof}. These methods have successfully captured complex structural distributions, expanded the accessible chemical and geometric design space, and increasingly supported property-directed MOF generation. These advances provide a strong foundation for exploring application-relevant pore-size regimes, where structural variety and precise aperture control are jointly valuable.

%Metal–organic frameworks (MOFs) offer extensive control over chemical composition and periodic geometry through their modular construction (Yaghi et al., 2003; Furukawa et al., 2013), making them promising candidates for gas storage, molecular separation, and catalysis (Furukawa et al., 2013). The combinatorial space of nodes, linkers, and network topologies, however, makes exhaustive exploration through synthesis or atomistic simulation impractical. Generative models address this challenge by learning from existing structures to propose new candidates (Yao et al., 2021). Recent approaches include coarse-grained diffusion in MOFDiff (Fu et al., 2024), torsion-aware flow matching in MOFFlow-2 (Kim et al., 2025a), conditional pore-field generation in MOFFUSION (Park et al., 2025), and physical-reward optimization in PRO-MOF (Liu et al., 2026). These developments expand the accessible structural design space and support property-directed generation, providing a foundation for controlling pore apertures while exploring distinct frameworks.

\begin{wrapfigure}{R}{0.5\textwidth}
\centering
\includegraphics[width=\linewidth]{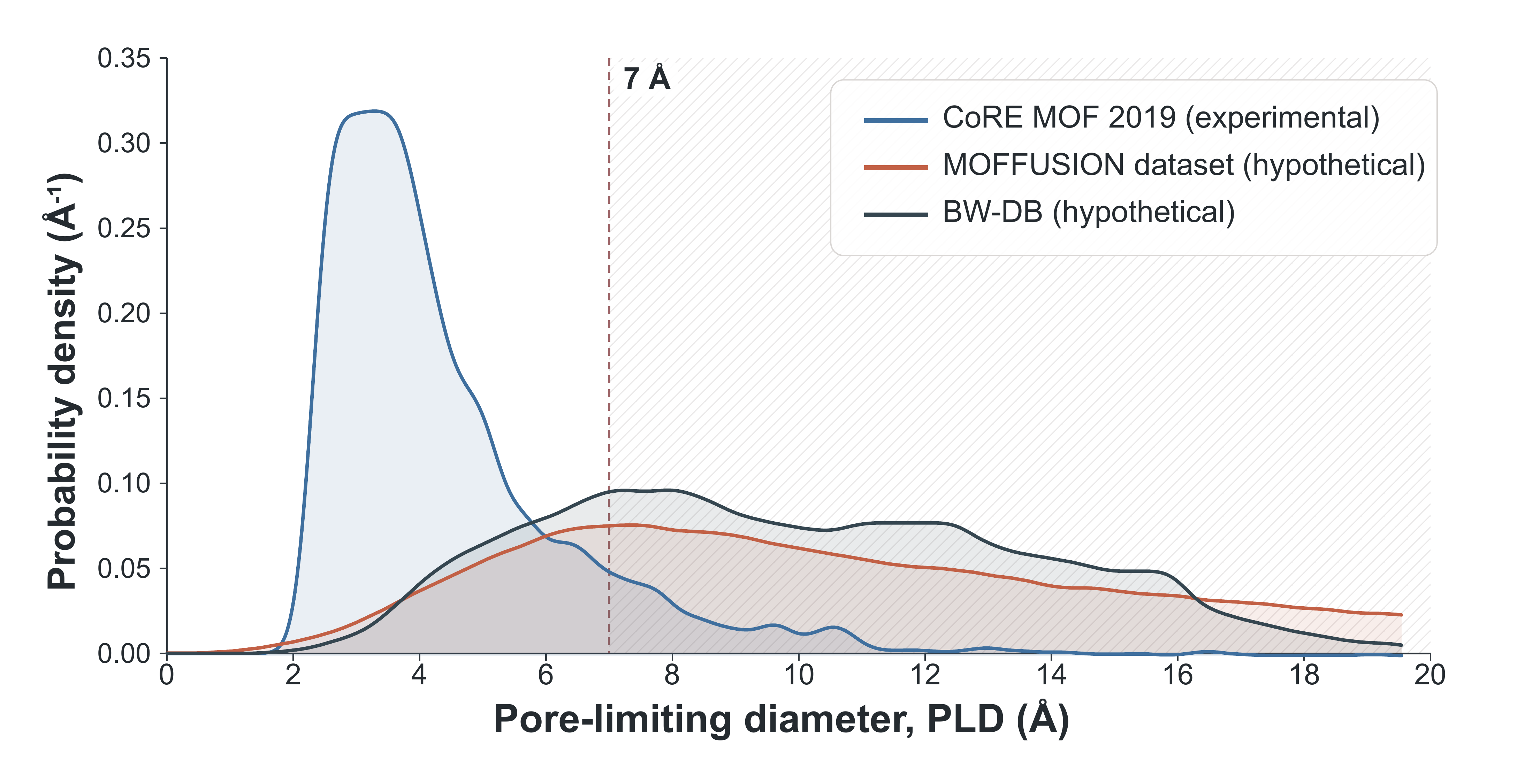}
\caption{\textbf{PLD coverage motivates target-directed generation.} Positive-PLD entries from selected CoRE MOF 2019 ASR/FSR records~\citep{chung2019core}, the processed MOFFUSION collection~\citep{park2025moffusion}, and BW-DB~\citep{boyd2019bwdb}. The dashed line marks 7~\AA.}
\label{fig:pld_motivation}
\end{wrapfigure}

Small pore apertures offer a particularly compelling focus within this broader design space. Figure~\ref{fig:pld_motivation} compares selected experimental and hypothetical collections through the pore-limiting diameter (PLD), the diameter of the largest spherical probe that can traverse a periodic pore network~\citep{willems2012zeo}. Among positive-PLD entries, 88.5\% of the selected CoRE collection lie below 7~\AA, compared with 21.7\% of the processed MOFFUSION collection and 23.56\% of the analyzed BW-DB collection~\citep{boyd2019bwdb}. BW-DB is particularly relevant here as an upstream source of training structures for MOFDiff, MOFFlow and MOFFlow-2~\citep{fu2024mofdiff,kim2025mofflow,kim2025mofflow2}. This consistent contrast across widely used hypothetical collections further strengthens the experimental motivation for small-aperture generation. Fine control is also especially valuable for molecular separation, where modest aperture changes can enable molecular sieving or alter the dominant separation mechanism~\citep{lin2018sieving,xue2018aperture}. Together, these distributional observations and application needs motivate dedicated generation within the small-aperture regime.

Generating MOFs within this regime is nevertheless demanding. A small PLD is determined by narrow bottlenecks along a connected periodic pore network, making aperture control particularly sensitive to the geometry surrounding these passages~\citep{willems2012zeo}. Local changes can appreciably narrow an opening or obstruct a previously traversable channel, so fine control requires preserving both the requested clearance and pore connectivity. In a pore-field-based approach, small-aperture training data and PLD conditioning guide exploration, while the requested clearance must ultimately be realized through discrete building-block selection and assembly. Finite-resolution fields efficiently capture pore morphology, while closely related linker chemistries can remain difficult to distinguish~\citep{park2025moffusion}. The challenge is therefore to realize narrowly specified apertures while retaining sufficient chemical and topological freedom to explore distinct frameworks. This motivates periodic global generation coupled to direct geometric verification and chemically feasible local correction.

To address these challenges, we introduce \textbf{CoFi-MOF}, a verifier-guided coarse-to-fine framework for pore-targeted MOF generation. Periodic U-DiT explores target-conditioned pore geometries while modeling relationships across unit-cell boundaries. A learned topology-aware Constructor maps each field to candidate building-block assignments through node prediction and conditional linker identification. Geometric measurements rank the assembled candidates by target agreement, and Multi-Edit-Former (MEF) proposes topology-preserving linker modifications for local refinement. Reassembly and evaluation determine whether an edit improves the aperture match, with the original candidates retained during final selection. We further expand PORMAKE’s carboxylate-based linker library and introduce \textbf{MOF-Edit-DB}, which pairs chemical transformations with measured framework-level pore responses.

Our main contributions are:
\begin{itemize}[leftmargin=3em, itemsep=1pt, topsep=2pt]
\item[\ding{182}] We propose \textbf{CoFi-MOF}, a coarse-to-fine system integrating periodic global generation, learned structural realization and measurement-driven local refinement for small-aperture MOF design. The contribution is the complete pipeline, with component-level controls used to identify the roles of its learned modules.

\item[\ding{183}] We develop a refinement procedure that selects assembled candidates by measured PLD and uses MEF to propose topology-preserving linker modifications. Reassembly and verification evaluate each modification while retaining the original candidates for final selection. Learned ranking improves first-edit accuracy over random editing and simpler learned rankers on matched task sets.

\item[\ding{184}] We expand PORMAKE's library with carboxylate-based linkers and build \textbf{MOF-Edit-DB}, a large-scale resource linking chemical edits to framework-level pore responses. The library and dataset support structural generation, property prediction and target-directed modification.

\item[\ding{185}] Under a fixed-upstream protocol with 25 measured realizations per attempt, CoFi-MOF achieves 86.20\% Hit@0.5 and 71.64\% Hit@0.2 over PLDs of 3--7~\AA, enriching chemically distinct candidates around each target.
\end{itemize}

\section{Related Work}

\paragraph{Property-directed MOF generation.}
CDVAE and DiffCSP encode crystal periodicity through boundary-aware representations and joint lattice--coordinate diffusion, respectively~\citep{xie2022cdvae,jiao2023diffcsp}. MOF generation has advanced from learned reticular representations to block-level diffusion and rigid-body or torsional flow matching~\citep{yao2021smvae,fu2024mofdiff,kim2025mofflow,kim2025mofflow2}. Property control uses conditional pore-field diffusion~\citep{park2025moffusion}, predictor-guided reinforcement learning for direct air capture~\citep{park2024rl}, and physical-reward optimization, including PLD targeting~\citep{liu2026promof}. Mofasa generates atoms, coordinates and lattices through all-atom latent diffusion~\citep{simkus2025mofasa}; EGMOF combines descriptor diffusion, Transformer decoding and predictor-based selection~\citep{han2025egmof}. CoFi-MOF couples toroidal attention over signed-distance fields with direct pore measurement, linking periodic proposals to verified aperture control.

\paragraph{From pore geometry to verified structures.}
Precise aperture control links continuous pore geometry to discrete chemistry. MOFFUSION enables conditional pore-field generation, while identifying finite resolution and ambiguity between similar-looking functional groups as reconstruction challenges~\citep{park2025moffusion}. Assembly realizes these components as periodic structures~\citep{pormake2021,park2024ghp}, whose narrow-channel clearances strongly affect molecular access~\citep{lin2018sieving,xue2018aperture}. Structural diversity is also an explicit optimization concern in PRO-MOF's reward design~\citep{liu2026promof}. CoFi-MOF connects periodic pore-field generation to learned, component-conditioned structural realization. Direct Zeo++ measurements guide aperture selection and assess local linker modifications~\citep{willems2012zeo}.

\paragraph{Linker optimization and edit supervision.}
Linker modification provides a route to tune pore chemistry within isostructural families~\citep{han2021linkersub}. Library augmentation and predictor-guided optimization expand the available design choices~\citep{yao2021smvae,cleeton2025dream}, while target-directed editing also benefits from supervision that links a specified chemical change to its realized pore response. MOF-Edit-DB provides these parent–edit pairs for training MEF. 
\section{Preliminaries}
\label{sec:preliminaries}

\paragraph{Pore-limiting and largest-cavity diameters.}
The internal geometry of a porous framework is commonly characterized by its pore-limiting diameter (PLD) and largest-cavity diameter (LCD). Let $\Omega(C)$ denote the void region of a framework $C$, and let $\rho_C(\mathbf r)$ be the radius of the largest sphere centered at $\mathbf r\in\Omega(C)$ that does not intersect the framework surface. These descriptors can be expressed as
\begin{equation}
    \operatorname{LCD}(C)
    =2\max_{\mathbf r\in\Omega(C)}\rho_C(\mathbf r),
    \qquad
    \operatorname{PLD}(C)
    =2\max_{\gamma\in\Gamma_{\mathrm{perc}}(C)}
      \min_{\mathbf r\in\gamma}\rho_C(\mathbf r),
\end{equation}
where $\Gamma_{\mathrm{perc}}(C)$ denotes the set of continuous paths that traverse the periodic pore network. LCD therefore measures the widest cavity that can be accommodated within the framework, whereas PLD measures the largest spherical probe that can pass through a connected periodic channel~\citep{willems2012zeo}. A framework may consequently contain spacious cavities while remaining accessible only through substantially narrower apertures. Because these bottlenecks directly govern molecular accessibility and sieving, we use PLD as the primary control variable and retain LCD as a complementary description of the internal pore volume. Both quantities are computed from realized structures using Zeo++.

Ultra-microporous MOFs have been described in terms of pore apertures below 5--7~\AA~\citep{adil2017ultramicroporous}. Here, we define the small-aperture regime by PLD below 7~\AA, rather than requiring every cavity to be smaller than this threshold.

\paragraph{Periodic representation of MOF pore geometry.}
A crystalline MOF is an infinite periodic structure specified by a lattice matrix $\mathbf A$ and the atomic or building-block arrangement within one unit cell. For a fractional coordinate $\mathbf u\in[0,1)^3$, its periodic images occupy
\begin{equation}
    \mathbf r_{\mathbf n}=\mathbf A(\mathbf u+\mathbf n),
    \qquad \mathbf n\in\mathbb Z^3.
\end{equation}
The unit-cell domain is therefore topologically equivalent to a three-dimensional torus: opposing cell faces are identified, and sites that appear distant in an isolated voxel grid may be immediate neighbors across a periodic boundary. Following MOFFUSION~\citep{park2025moffusion}, we describe pore morphology using multichannel signed-distance fields (SDFs) sampled over this periodic domain, with distances evaluated against the periodically replicated framework surface. This representation compactly captures continuous pore geometry and cross-boundary connectivity, but it does not uniquely specify the discrete topology, nodes, or linkers needed to construct a chemically explicit MOF. Generating a periodic pore field and realizing it as a discrete framework therefore remain two closely connected but fundamentally distinct problems.
\section{Method}
\label{sec:method}

% CoFi-MOF combines global exploration of periodic pore geometry with local correction of realized apertures. Periodic U-DiT generates target-conditioned fields in the fixed signed-distance-field latent space of a pretrained VQ-VAE. A topology-aware Constructor maps each field to discrete building-block assignments, which are assembled and measured with Zeo++. MEF then proposes linker modifications to a promising editable candidate, and final selection uses the measured target error across the original and edited structures. 

\begin{figure}[t]
\centering\includegraphics[width=\linewidth]{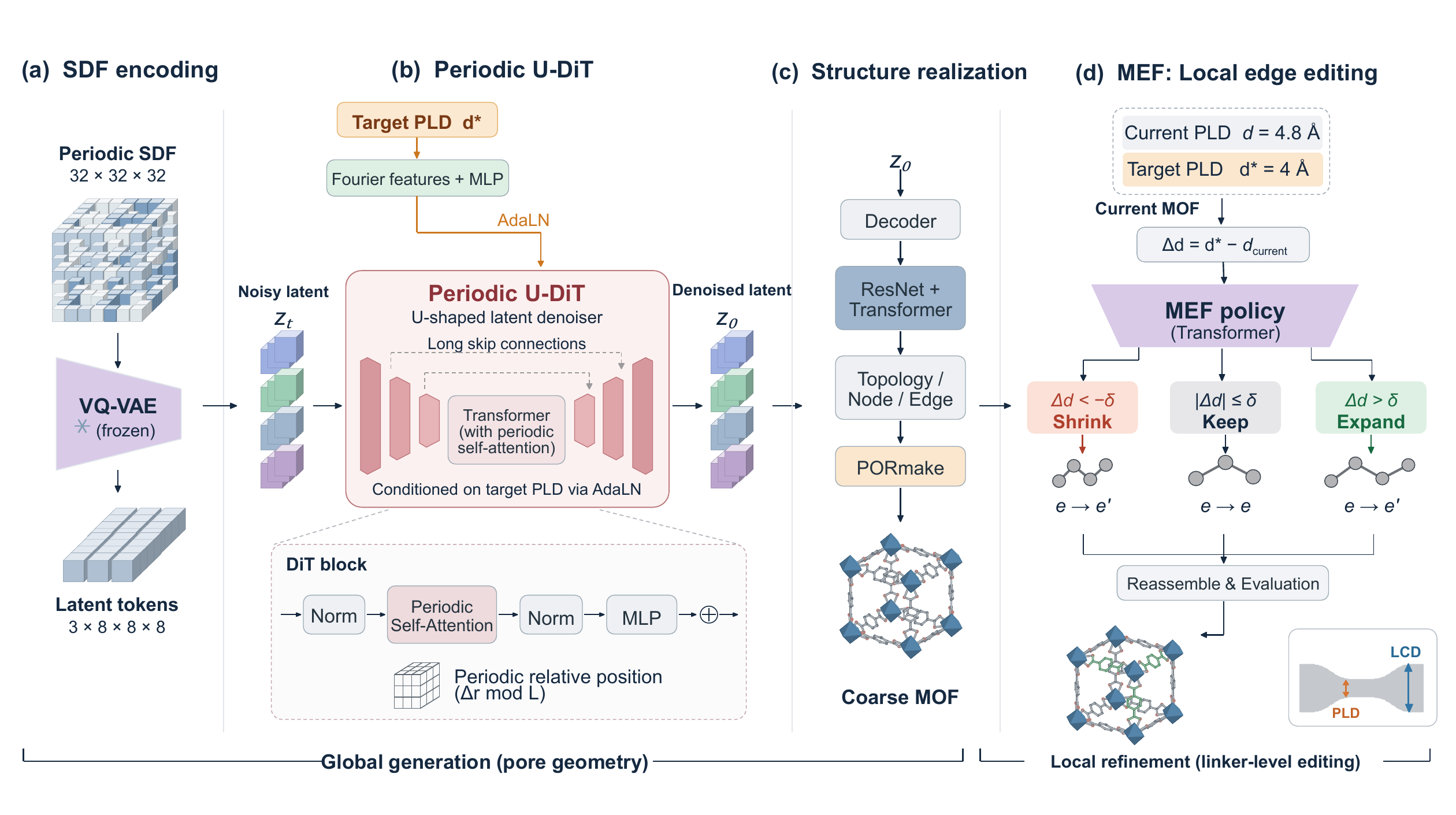}\caption{\textbf{Coarse-to-fine generation in CoFi-MOF.} (a) A frozen VQ-VAE encodes periodic signed-distance fields; (b) Periodic U-DiT generates target-conditioned pore geometry; (c) the learned Constructor realizes each field through node and conditional linker prediction; and (d) Zeo++ verification and MEF linker editing refine structures toward the requested PLD. Panel (d) illustrates pore contraction, retention and expansion; MEF predicts explicit chemical modifications. Algorithm~\ref{alg:cofi} specifies the complete generation and single-batch refinement procedure.}\label{fig:pipeline}\end{figure}

CoFi-MOF concentrates candidate MOFs around a requested aperture by translating the target pore size into diverse discrete realizations while exploring distinct chemistries and topologies. It does so in three stages: (1) PLD-conditioned Periodic U-DiT generates pore geometries with cross-boundary spatial modeling; (2) a topology-aware Constructor maps these geometries to candidate frameworks via node prediction and conditional linker identification; (3) geometric verification and MEF-based refinement apply geometric measurements to select promising assemblies and target-directed linker edits for finer aperture control. Together, these stages link global exploration of periodic pore geometry with local chemical refinement. Figure~\ref{fig:pipeline} illustrates the overall workflow.

\subsection{PLD-conditioned periodic MOF generation}
\paragraph{Periodic SDF latent space.}
The global stage must increase access to small-aperture pore geometries without restricting generation to a narrowly filtered structural prior. Following MOFFUSION~\citep{park2025moffusion}, we represent each MOF by a four-channel periodic signed-distance field $x\in\mathbb R^{4\times32\times32\times32}$ and use a frozen VQ-VAE~\citep{oord2017vqvae} to obtain an unquantized continuous latent $z_e\in\mathbb R^{3\times8\times8\times8}$. Diffusion~\citep{ho2020ddpm,rombach2022ldm} is trained directly on $z_e$, without codebook quantization or latent normalization. At generation time, the sampled latent is quantized once and decoded as $\widehat x=D(Q(\widehat z_e))$. This representation retains continuous pore morphology for global generation while deferring discrete chemical realization to the Constructor.

\paragraph{Periodic U-DiT.}
Because MOF pore channels extend across unit-cell boundaries, the latent denoiser must model long-range geometry without introducing artificial cell edges. We train an $\epsilon$-prediction denoiser on $z_s=\sqrt{\bar\alpha_s}z_e+\sqrt{1-\bar\alpha_s}\epsilon$:
\begin{equation}
 \mathcal L_{\mathrm{diff}}=\mathbb E\big[\|\epsilon-\epsilon_\theta(z_s,s,d^\star)\|_2^2\big].
\end{equation}
The $8^3$ latent grid is treated as 512 tokens and processed by eight constant-resolution Transformer blocks with long feature skips~\citep{bao2023uvit}. To make global attention consistent with the periodic grid, head $h$ assigns sites $r_i,r_j$ the attention logit
\begin{equation}
 a_{ij}^{(h)}=\langle q_i^{(h)},k_j^{(h)}\rangle/\sqrt{d_h}
 +b_h\big((r_i-r_j)\bmod8\big).
\end{equation}
The learned bias represents directed displacement on a discrete torus, including neighbors across opposite cell faces. No absolute spatial embedding is used. A Fourier encoding of $d^\star$ enters every block through adaptive layer normalization~\citep{peebles2023dit}. Spatially shared conditioning and token-wise operations preserve equivariance to integer cyclic shifts. Deterministic DDIM sampling~\citep{song2021ddim} produces $\widehat z_e$, which the frozen quantizer and decoder map to a target-conditioned periodic pore field. Detailed architecture and numerical probes are provided in Appendices~\ref{app:periodic_architecture} and~\ref{app:periodic_audit}.

\subsection{Learned topology-aware structure realization}
The Constructor completes the transition from periodic pore geometry to discrete chemistry. Given a generated field and its predicted topology, a Node Transformer identifies inorganic and optional organic nodes. An Edge Transformer then conditions on both the field and the proposed nodes to identify compatible linkers, explicitly accounting for component dependencies during structural realization~\citep{vaswani2017attention}. The models rank alternative building-block assignments, which PORMAKE assembles into periodic frameworks~\citep{pormake2021}. These alternatives are chemical realizations of the same generated field. Zeo++ subsequently determines their agreement with the target aperture.

\paragraph{One-to-many chemical realization.}
A finite-resolution pore field is a coarse geometric blueprint rather than a unique specification of the underlying chemical structure. Distinct building-block assignments can be consistent with similar pore morphologies while producing different atomic-scale apertures~\citep{park2025moffusion}. Committing to a single highest-ranked assignment may therefore discard alternative realizations that better satisfy the target PLD. Instead of repeatedly sampling the global diffusion model, CoFi-MOF explores multiple plausible chemical realizations conditioned on the same generated field: generate geometry once, realize chemistry multiple times, and refine locally. This decoupling amortizes the cost of global geometry generation across candidate structures and concentrates subsequent exploration in the discrete chemical space, where measurement feedback resolves the target-dependent differences among realizations.

We expand PORMAKE's building-block library with carboxylate-based linkers informed by reviews of Zr-MOFs and titanium coordination polymers~\citep{bai2016zrmofs,du2024titanium}. This extension supports a Constructor vocabulary of more than 600 linker classes and is packaged with the MOF-Edit-DB resources. Appendix~\ref{app:constructor_architecture} describes the learned heads and their evaluation; Appendix~\ref{app:dataset_scope} details library construction.

\subsection{Staged verification and MEF refinement}
\paragraph{Measurement-based candidate selection.}
The global branch broadens access to small-aperture candidates, whereas precise target control is determined from their realized pore geometry. Learned scores prioritize which recipes and edits to evaluate; Zeo++ then verifies their actual pore apertures at both realization stages. We use this fixed geometric verifier~\citep{willems2012zeo} and define

\begin{equation}
 V(C)=\operatorname{PLD}_{\mathrm{Zeo++}}(C),\qquad
 e(C;d^\star)=|V(C)-d^\star|.
\end{equation}
Here, Zeo++ measures each assembled MOF. Verification takes approximately 0.3~s per structure in our CPU benchmark, making geometric measurement practical throughout structure realization and local refinement (Appendix~\ref{app:verifier_runtime}).

Let $\mathcal C$ contain the Constructor's successfully assembled structures with finite positive PLD. The closest initial match is
\begin{equation}
 C_0=\operatorname*{arg\,min}_{C\in\mathcal C}e(C;d^\star).
\end{equation}
If no candidate has a valid measurement, generation is unsuccessful. Otherwise, refinement starts from the closest match whose linker supports the permitted chemical modifications. This source $C_s$ can differ from $C_0$, which remains available for final selection. MEF receives the source's linker state, measured PLD, and target in a single call, including when the source already satisfies the reporting tolerance. This permits further improvement at finer pore-size resolution.

\begin{wrapfigure}{r}{0.5\textwidth}
\centering
\includegraphics[width=\linewidth]{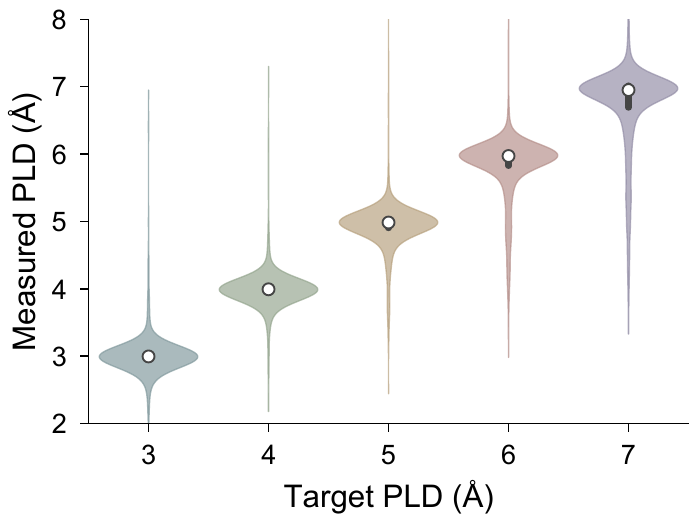}
\caption{\textbf{Target-conditioned PLD distributions.} CoFi-MOF at 3--7~\AA. Circles denote medians; bars, interquartile ranges. Widths show Gaussian KDEs (bandwidth 0.15~\AA) using all observations, displayed over 2--8~\AA.}
\label{fig:pld_distribution}
\end{wrapfigure}
\paragraph{Target-directed linker correction.}
MEF broadens local chemical exploration beyond the Constructor's initial assignments by ranking complete destination assignments over at most three selected linker sites. It conditions on the current chemical states, invariant linker geometry, measured PLD $d_0$, and requested target $t$. The difference $t-d_0$ provides the required correction direction and magnitude. For a legal assignment $a$ and source state $s$, its score is
\begin{equation}
 S_\phi(a)=\sum_i u_\phi(i,a_i,t)
 -\lambda_{\mathrm{edit}}d_H(a,s),
\end{equation}
where $d_H$ penalizes unnecessary changes. Deterministic chemistry and constructability masks remove unsupported assignments before ranking. MEF is trained on measured parent--edit relationships from MOF-Edit-DB using a multi-positive listwise objective; crucially, unmeasured assignments are not treated as negatives. MEF data are partitioned by linker lineage, with disjoint base-linker identities across training, development and held-out sets (Appendix~\ref{app:mef_split}). During materialization and reassembly, the surrounding topology and node identities remain fixed. Architecture, supervision, and the frozen-model scope are detailed in Appendix~\ref{app:mef_architecture}, with dataset construction described in Appendix~\ref{app:dataset}.

\paragraph{Verified pore refinement.}
MEF proposes up to $M$ distinct linker modifications according to $S_\phi$, excluding assignments that leave the source unchanged. Each proposal is reassembled with the original topology and nodes, and Zeo++ measures its resulting aperture. The learned score thus guides chemical exploration, while the measured pore response determines which modification, if any, should be retained. Let $\mathcal F$ contain all original and edited structures with finite positive measured PLD. We select
\begin{equation}
 C_{\mathrm{sel}}=\operatorname*{arg\,min}_{C\in\mathcal F}e(C;d^\star).
\end{equation}
Retaining the original structures ensures that unsuccessful edits cannot displace a better initial match; ties favor the original candidate. If refinement produces no eligible alternative, $C_0$ is returned.

Each attempt begins with one generated pore field and returns at most one MOF after learned structural realization, geometric verification and local refinement. Both the generator and MEF accept continuous PLD conditions; the reported experiments use an integer target grid. Algorithm~\ref{alg:cofi} in Appendix~\ref{app:verifier_protocol} provides the complete inference procedure.

\section{Experiments}
\label{sec:experiments}
We evaluate small-aperture generation, the contributions of periodic modeling and geometric verification, and learned linker refinement. Component-level studies assess Periodic U-DiT and the Constructor separately, while the main ablations examine their integration with verification and MEF. The primary benchmark uses PLDs measured on assembled structures with Zeo++.

\subsection{PLD control benchmark}
\label{sec:pld_benchmark}
The main benchmark tests whether target conditioning and verified refinement concentrate the final pore distribution around each requested aperture.

We consider target PLDs of $\{3,4,5,6,7\}$~\AA, spanning the small-aperture range that motivates this work. Each neural pipeline receives 1{,}000 fixed upstream proposals per target and 25 measured structural realizations per proposal, with the candidate closest to the target retained. CoFi-MOF additionally evaluates one MEF edit against the verified pool. Table~\ref{tab:main} compares the resulting outputs with target-adapted GA, MOFDiff, MOFFlow-2 and MOFFUSION~\citep{fu2024mofdiff,kim2025mofflow2,park2025moffusion}. MOFDiff ranks a posterior latent pool by predicted target agreement, MOFFUSION assembles its 25 highest-ranked building-block assignments under the predicted topology, and MOFFlow-2 resamples the three-dimensional realization of a fixed building-block sequence. GA uses a separate 1{,}000-proposal search protocol. Model configurations, structural realization and data sources are specified in Appendices~\ref{app:physical_eval_protocol}, \ref{app:baseline_settings} and~\ref{app:budget25}.

\begin{table}[t]
\centering\small
\setlength{\tabcolsep}{2.5pt}
\caption{\textbf{Pore-targeted MOF generation across PLD targets of 3--7~\AA.} The four neural pipelines receive 1{,}000 fixed upstream proposals per target and 25 measured structural realizations per proposal, with the closest assembled PLD retained. CoFi-MOF additionally applies one MEF edit to the verified selection and compares it with the original pool. GA is a separate surrogate-scored search baseline with 1{,}000 proposals per target. The protocols differ in training distributions and realization spaces. Metrics use the 1{,}000 allocated attempts per target, with assembly and measurement failures counted as misses; MAE and signed bias (\AA) use measurable outputs on assembled structures. Best value per target and metric is bolded.}
\label{tab:main}
\resizebox{\textwidth}{!}{%
\begin{tabular}{llccccc}
\toprule
Method & Metric & 3 & 4 & 5 & 6 & 7\\
\midrule
GA & Yield & 88.40 & 92.00 & 88.80 & 92.40 & 90.90\\
 & Hit@0.2/0.5 & 36.10 / 59.10 & 59.90 / 71.10 & 64.00 / 67.20 & 71.40 / 72.40 & 59.00 / 72.00\\
 & MAE/bias & 1.270 / +1.135 & 1.193 / +1.155 & 0.920 / +0.754 & 1.396 / +1.164 & 0.954 / +0.625\\
\midrule
MOFDiff + selection & Yield & 91.10 & 92.50 & 93.10 & 93.60 & 91.80\\
 & Hit@0.2/0.5 & 50.00 / 63.10 & 61.70 / 74.30 & 45.50 / 59.60 & 33.60 / 45.00 & 22.60 / 35.40\\
 & MAE/bias & 0.530 / +0.480 & 0.328 / +0.112 & 0.514 / $-0.288$ & 0.731 / $-0.553$ & 1.018 / $-0.892$\\
\midrule
MOFFlow-2 (cond.) & Yield & 91.20 & 91.60 & 91.60 & 91.60 & 91.60\\
 & Hit@0.2/0.5 & 42.40 / 54.30 & 66.20 / 75.30 & 58.10 / 72.30 & 33.40 / 44.90 & 16.70 / 23.30\\
 & MAE/bias & 0.661 / +0.639 & 0.287 / +0.182 & 0.322 / $-0.119$ & 0.751 / $-0.662$ & 1.455 / $-1.396$\\
\midrule
MOFFUSION (PLD) & Yield & 94.80 & 94.30 & 96.30 & 96.70 & 97.20\\
 & Hit@0.2/0.5 & 58.40 / 72.80 & 57.40 / 72.50 & 59.30 / 76.00 & 57.00 / 74.90 & 58.40 / 74.30\\
 & MAE/bias & 0.435 / +0.327 & 0.426 / +0.284 & 0.375 / +0.168 & 0.402 / \textbf{+0.035} & \textbf{0.406} / \textbf{$-0.072$}\\
\midrule
\textbf{CoFi-MOF} & \textbf{Yield} & \textbf{100.00} & \textbf{100.00} & \textbf{100.00} & \textbf{99.90} & \textbf{99.50}\\
 & \textbf{Hit@0.2/0.5} & \textbf{78.50 / 91.20} & \textbf{79.90 / 91.70} & \textbf{73.90 / 89.60} & \textbf{67.97 / 82.08} & \textbf{58.29 / 76.88}\\
 & \textbf{MAE/bias} & \textbf{0.176 / +0.042} & \textbf{0.183 / +0.014} & \textbf{0.204 / $-0.045$} & \textbf{0.300} / $-0.186$ & 0.424 / $-0.276$\\
\bottomrule
\end{tabular}%
}
\end{table}

\revised{\paragraph{Training distributions and realization budgets.} Table~\ref{tab:main} compares complete target-adapted pipelines under a matched budget of upstream proposals and measured realizations (Appendix~\ref{app:training_compatibility}).}

\paragraph{Target control and success.}
Let $N_{d^\star}$ be the number of allocated attempts at target $d^\star$, and let $\mathcal V_{d^\star}$ contain attempts whose selected output has a finite positive measured PLD. Measurable yield is $|\mathcal V_{d^\star}|/N_{d^\star}$. The target hit rate is
\begin{equation}
 H_\delta(d^\star)=\frac{1}{N_{d^\star}}\sum_{i\in\mathcal V_{d^\star}}
 \mathbf{1}\!\left[|d_i-d^\star|\leq\delta\right].
\end{equation}
Hit@0.5 is the primary metric; Hit@0.2 and Hit@1.0 assess tighter and broader tolerances. Assembly and measurement failures count as misses. MAE and signed bias, $d_i-d^\star$, use measurable outputs. Pooled results aggregate counts and errors across targets. Yield measures pore measurability, not synthetic feasibility.

\paragraph{Accurate pore realization across targets.}
Figure~\ref{fig:pld_distribution} shows strong target concentration. \revised{Under the shared 25-candidate budget, CoFi-MOF achieves 86.20\% Hit@0.5 and 71.64\% Hit@0.2, with 99.88\% measurable yield and 0.257~\AA\ MAE. Control is particularly strong at smaller apertures, with Hit@0.5 of 91.20\% and 91.70\% at 3 and 4~\AA.} The evaluated neural baselines reach at most \revised{76.00\%} Hit@0.5 and \revised{66.20\%} Hit@0.2 across targets, with measurable yields of \revised{91.10--97.20\%}. GA reaches 59.10--72.40\% attempt-level Hit@0.5, yet 74.7--78.7\% of its proposals repeat a topology--node--linker combination within a target; CoFi-MOF keeps 98.8\% of measured outputs in the initial batch structurally distinct (Appendix~\ref{app:rigid_results}). CoFi-MOF thus combines higher target agreement with higher measurable yield under the evaluated protocols. The resulting candidate pool also serves downstream screening: a force-field study of 500 tasks reaches 78.2\% Hit@0.5 after relaxation and shortlist selection (Appendix~\ref{app:relaxed_realizations}).

\revised{Beyond within-batch uniqueness, 94.93\% of the evaluated outputs have recipe combinations absent from the global corpus. No crystal matches are found against the evaluated global corpus, reconstructed CoRE subset and BW-DB collection under the specified matching protocol (Appendix~\ref{app:structural_novelty}).}

\subsection{\mbox{Periodic generation and structural realization}}
\label{sec:periodic_results}

To assess generation independently of candidate selection and editing, we compare unconditional Periodic U-DiT with the MOFFUSION 3D U-Net using matched data, a frozen VQ-VAE, 60,000 training updates and 500 shared initial noises. Periodic U-DiT increases assembly success from 64.0\% to 71.2\%. Across 7,500 cyclic-shift probes, median relative equivariance error falls from $2.859\times10^{-2}$ to $1.156\times10^{-7}$. \revised{End-to-end validity after assembly, relaxation and Zeo++ characterization rises from 58.0\% to 67.4\% ($+9.4$ points; one-sided exact McNemar $p=8.2\times10^{-4}$), and the denoiser uses $6.5\times$ fewer parameters than the 3D U-Net at the same training budget.} Appendix~\ref{app:periodic_audit} provides the protocols and detailed diagnostics, and Appendix~\ref{app:constructor_architecture} reports the complementary Constructor studies.

\subsection{Contributions of verification and linker refinement}
\label{sec:ablations}
Table~\ref{tab:ablations} compares four configurations on the same generated pore fields. The MEF top-1 row is the fixed-upstream configuration of Table~\ref{tab:main}. The MEF top-10 row spends up to nine additional measured candidates per attempt; since no baseline receives that budget, we report it here as an ablation rather than in the main comparison. CoFi-MOF w/o verifier and MEF directly returns the Constructor's globally highest-ranked assignment under the joint node--linker score, without measured candidate selection or local editing. CoFi-MOF w/o MEF adds Zeo++ selection among the Constructor's realized alternatives. Both CoFi-MOF configurations further apply MEF and re-verify its edits. Zeo++ characterizes all final outputs for evaluation, including the direct-output variant.

\begin{table}[t]
\centering%\small
\caption{\textbf{Contributions of geometric verification and MEF.} All variants retain Periodic U-DiT and the learned Constructor and share the 5{,}000 allocated attempts at targets 3--7~\AA. Without verification and MEF, the model directly returns the assignment with the highest joint node--linker score. Hit rates use all allocated attempts, with assembly and measurement failures counted as misses; MAE (\AA) uses measurable outputs.}
\label{tab:ablations}
\begin{tabular}{lrrr}
\toprule
Variant & Hit@0.2 & Hit@0.5 & MAE\\
\midrule
w/o verifier and MEF & 7.62 & 17.54 & 1.734\\
w/o MEF & 68.24 & 83.34 & 0.303\\
CoFi-MOF (MEF top-1) & 71.64 & 86.20 & 0.257\\
CoFi-MOF (MEF top-10) & 77.46 & 88.10 & 0.220\\
\bottomrule
\end{tabular}
\end{table}

\paragraph{Verification completes pore-targeted realization.}
Combining the learned Constructor with geometric verification raises Hit@0.5 from 17.54\% to 83.34\% before local editing, while MAE decreases from 1.734 to 0.303~\AA. \revised{This verified stage exceeds every baseline at 3--6~\AA{} and is comparable at 7~\AA{} (73.30\% versus 74.30\%).} This improvement reflects the complementary roles of global pore-field generation and discrete chemical realization. Periodic U-DiT provides a target-conditioned geometric blueprint, narrowing subsequent exploration toward the desired pore regime. Rather than repeatedly generating new pore fields, the learned Constructor explores multiple plausible chemical realizations of the same blueprint. Geometric verification resolves their differences in realized aperture, while MEF further reduces residual target errors through local chemical modifications. This hierarchical allocation of exploration enables pore control at progressively finer levels of structural resolution.

\paragraph{Local editing further sharpens pore control.}
\revised{Local refinement with a single verified edit raises Hit@0.5 from 83.34\% to 86.20\%, adding 143 successful structures}, while MAE falls from 0.303 to 0.257~\AA. The gain is particularly pronounced at the tighter tolerance. Hit@0.2 increases from 68.24\% to 71.64\%, adding 170 precise matches. Allowing up to ten edits reaches 88.10\% Hit@0.5 and 77.46\% Hit@0.2 (Table~\ref{tab:ablations}), at the cost of nine further measured candidates per attempt, which is why we keep it out of the fixed-upstream comparison. These results establish local chemical refinement as an effective complement to verified structural realization.

Hit@0.5 improves at all five targets and in both sampling batches, with original matches retained. The pooled signed error shifts from $-0.036$ to $-0.111$~\AA, indicating a modest downward bias alongside improved absolute accuracy. Target-wise results and paired uncertainty estimates are reported in Appendix~\ref{app:rigid_results}.

\subsection{Learned linker editing}
\label{sec:mef_control}
\begin{figure}[t]
\centering
\begin{minipage}[b]{0.49\linewidth}\centering
\includegraphics[width=\linewidth]{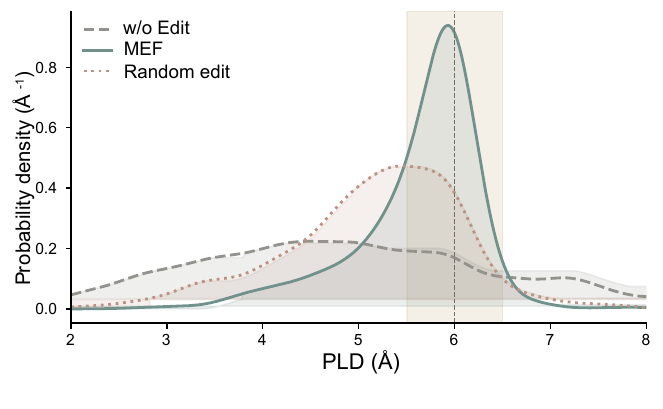}\\[-2pt]\small (a) First-proposal pore distributions
\end{minipage}\hfill
\begin{minipage}[b]{0.49\linewidth}\centering
\includegraphics[width=\linewidth]{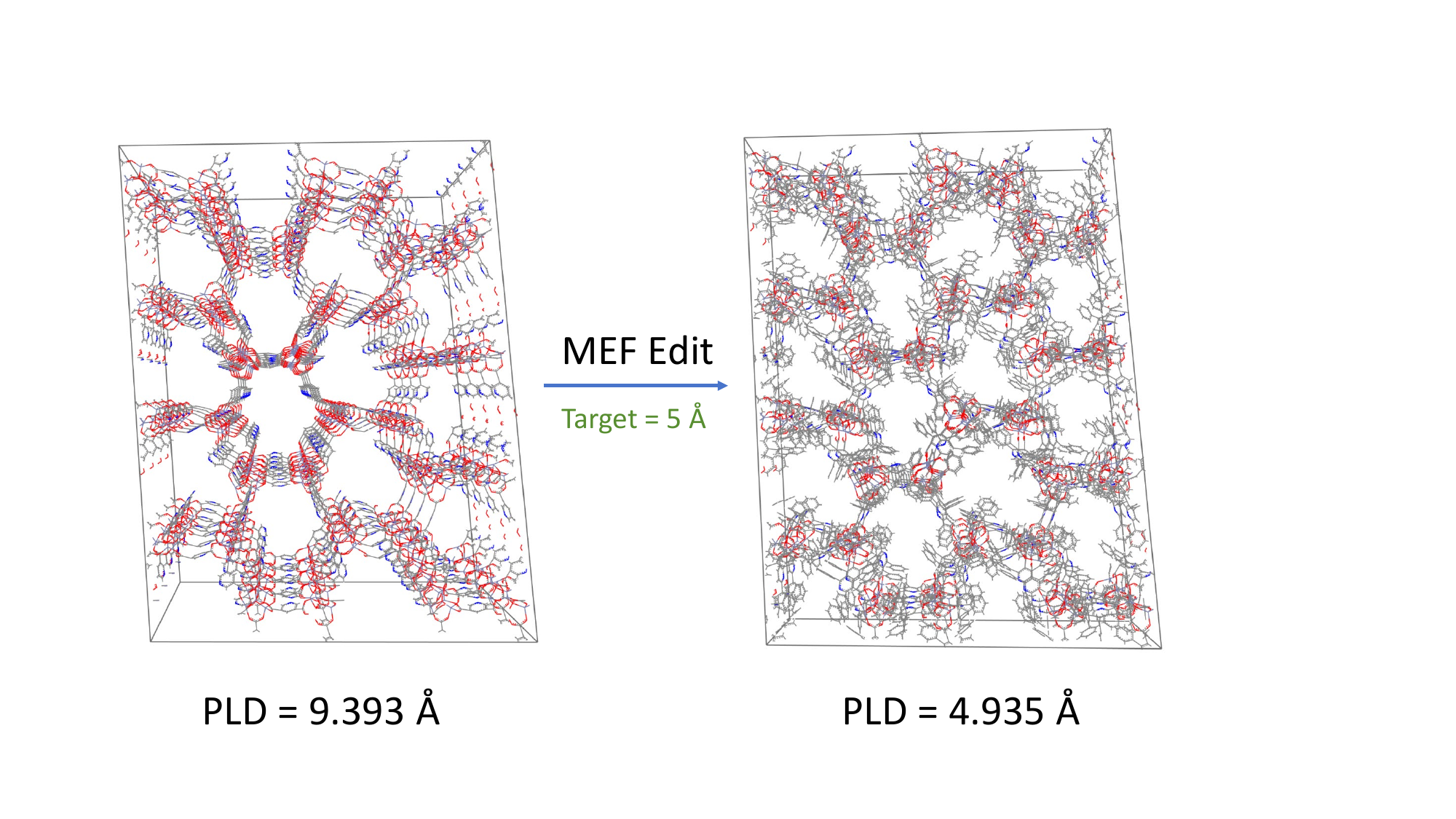}\\[-2pt]\small (b) Target-directed linker editing
\end{minipage}
\caption{\textbf{Learned edits concentrate pore apertures near the target.} (a) Measured PLDs at 6~\AA\ for MEF and random first proposals from shared editable sources, with the direct output of the same pipeline (w/o verifier and MEF) as reference. Curves use 890 jointly measurable tasks; reported hit rates use all 937. (b) A representative edit brings PLD from 9.393 to 4.935~\AA\ toward a 5~\AA\ target. Further examples appear in Appendix~\ref{app:edit_cases}.}
\label{fig:mef_control}
\end{figure}

To isolate the quality of learned edit proposals, we compare MEF with random editing on 937 paired tasks at a target PLD of 6~\AA. Both editors act on the same selected source for each task, and their first proposals are evaluated without post-edit candidate selection. MEF achieves 68.20\% Hit@0.5, substantially exceeding random editing at 33.62\%, with a paired gain of 34.58 percentage points (95\% bootstrap CI [31.38, 37.89]). MAE is also markedly lower, at 0.484 versus 1.004~\AA. Figure~\ref{fig:mef_control} connects this sharper target concentration to a representative structural edit. All 937 tasks, including failed proposals, enter the hit-rate denominator; full results are given in Appendix~\ref{app:mef_control}. \revised{MEF also outperforms three learned ranking baselines trained on the same examples, reaching 70.59\% Top-1 Hit@0.5 versus 39.39\% for the strongest XGBoost baseline on 1,129 matched tasks (Appendix~\ref{app:ranking_baselines}).}

\subsection{MOF-Edit-DB and an expanded linker library}
\label{sec:dataset}
MOF-Edit-DB packages an expanded PORMAKE library of carboxylate-based linkers with a large-scale framework edit-response resource. The library directly extends the building blocks available to PORMAKE and supports both learned structural realization and chemically specified linker modification. It contains 138,771 force-field-relaxed framework records, of which 52.5\% have PLDs below 7~\AA, and 108,145 matched parent--edit comparisons. Construction links chemically specified linker modifications to complete periodic frameworks and their relaxed pore responses, with systematic MOFChecker-based screening and record-level quality annotations~\citep{jin2025mofchecker}. These linked representations support pore-property prediction, target-conditioned edit ranking and small-aperture generative modeling, extending the resource's value beyond MEF training. Appendix~\ref{app:dataset} details library construction, quality annotations and data access; additional editing examples appear in Appendix~\ref{app:edit_cases}. The data package and access documentation are described in the supplementary materials.

\section{Limitations}
\label{sec:limitations}
CoFi-MOF operates within the topologies, building-block libraries and assembly geometries supported by PORMAKE. The current pipeline does not explicitly optimize axial orientations of two-connected linkers. Orientation-aware placement and task-specific physical validation remain important extensions.

\section{Conclusion}
CoFi-MOF connects periodic pore-field generation, learned structural realization and verified linker editing for target-directed MOF design. Across PLD targets of 3--7~\AA, geometric realization achieves 83.34\% Hit@0.5, and one verified MEF edit raises this to 86.20\% under the fixed-upstream protocol. By concentrating verified, chemically distinct candidates near each requested aperture, the system supplies actionable shortlists for downstream simulation and experimental selection. The expanded PORMAKE linker library and MOF-Edit-DB complement this system with resources for learning and evaluating chemical edits and their framework-level pore responses.

\clearpage
\section*{Data Availability}
The appendix records the data sources, implementation details and evaluation protocols. MOF-Edit-DB includes data tables, edit catalogs, quality annotations and a structure manifest; geometry availability is documented in the accompanying README.

\section*{Ethics and Scope}
This computational study involves no human subjects, personal data or animal experimentation. The generated frameworks are hypothetical and evaluated in silico. The reported measurements characterize pore geometry under the stated assembly and Zeo++ protocols; they do not establish chemical stability, synthesizability or experimental performance.

\section*{AI-use Statement}
We used LLMs to improve the readability and fluency of the manuscript. They did not contribute to the scientific ideas or algorithm design. All substantive content, reasoning and conclusions were developed by the authors.

% \clearpage
\bibliographystyle{iclr2027_conference}
\bibliography{references}
\clearpage
\appendix
\section{Model architecture and training}
\label{app:implementation}
CoFi-MOF separates periodic pore-field generation from discrete framework realization and local linker refinement. This section details the representations and learned components; Appendix~\ref{app:verifier_protocol} specifies how they interact during inference.

\subsection{Periodic pore representation and structural realization}
\label{app:representation}
Each MOF is represented by a four-channel $32^3$ signed-distance field and encoded into a continuous latent of shape $3\times8^3$. The pretrained VQ-VAE remains fixed during diffusion training. The denoiser operates on unquantized encoder latents; sampled latents are quantized once before decoding into pore fields.

The learned Constructor converts each decoded field into discrete building-block assignments. PORMAKE realizes these assignments as periodic atomic structures, and Zeo++ measures their pore apertures. The component models and structural realization procedure are detailed below.

\subsection{Periodic U-DiT architecture}
\label{app:periodic_architecture}
\paragraph{Tokenization and network layout.}
The denoiser receives a noisy encoder latent $z_t\in\mathbb{R}^{3\times8\times8\times8}$. Each spatial site becomes one three-channel token, giving 512 tokens; a shared linear projection maps each token to width 768. All eight transformer blocks retain this resolution and width. Each block has 12 attention heads of dimension 64 and a feed-forward expansion ratio of four. The U-shaped layout refers to long feature skips across depth, with no spatial downsampling or upsampling. A token-wise output head predicts three noise channels, which are reshaped to the input lattice. The unconditional backbone has 95,706,627 parameters; continuous PLD conditioning extends the same architecture.

\paragraph{Toroidal relative attention.}
For sites $r_i,r_j\in\{0,\ldots,7\}^3$, define the directed wrapped displacement $\Delta_{ij}=(r_i-r_j)\bmod8$. The lookup index is $64\Delta_{ij,x}+8\Delta_{ij,y}+\Delta_{ij,z}$. Each block learns its own bias table $b\in\mathbb{R}^{12\times512}$, with one scalar per head and wrapped displacement. A shared linear map produces queries, keys and values. For head $h$,
\begin{equation}
 A_{ij}^{(h)}=\operatorname{softmax}_{j}\!\left(
 \frac{\langle q_i^{(h)},k_j^{(h)}\rangle}{\sqrt{64}}+b_h[\Delta_{ij}]
 \right),\qquad
 o_i^{(h)}=\sum_j A_{ij}^{(h)}v_j^{(h)}.
\end{equation}
Attention spans all 512 sites. Head outputs are concatenated and linearly projected to width 768. Directed offsets retain orientation while identifying displacements across opposite cell faces. The model has no absolute spatial embedding; the sinusoidal timestep embedding encodes diffusion time only. Bias tables are initialized to zero and learned independently across blocks.

\paragraph{Conditioning and AdaLN-Zero blocks.}
A 768-dimensional sinusoidal timestep embedding, with frequency base 10,000, passes through a $768\!\to\!3072\!\to\!768$ MLP with SiLU activation to obtain $c_t$. The unconditional model uses $c=c_t$. The PLD-conditioned model first normalizes its PLD target as $u=d^\star/15$ and constructs
\begin{equation}
 \phi(d^\star)=\big[u,\sin(2\pi u),\cos(2\pi u),
 \sin(4\pi u),\cos(4\pi u),\sin(8\pi u),\cos(8\pi u)\big].
\end{equation}
A $7\!\to\!128\!\to\!128$ MLP uses SiLU after each layer; a linear $128\!\to\!768$ projection gives $c_d$. The conditioned model uses $c=c_t+c_d$ in every block and the output head. The PLD projection is zero-initialized, so adding the conditioner initially preserves the unconditional computation. The implemented target domain is $[0,15]$~\AA. Conditioning is spatially broadcast through adaptive normalization, with no conditioning dropout or classifier-free guidance. The conditioner adds $116{,}608$ parameters, giving $95{,}823{,}235$ trainable parameters in total.

Each block maps $c$ through SiLU and a linear $768\!\to\!4608$ layer to obtain attention and MLP shift, scale and gate vectors $(\beta_a,\gamma_a,g_a,\beta_m,\gamma_m,g_m)$. For input $H$, the block computes
\begin{align}
 \widetilde H &= H+g_a\odot\operatorname{Attn}
 \big((1+\gamma_a)\odot\operatorname{LN}_a(H)+\beta_a\big),\\
 H' &=\widetilde H+g_m\odot\operatorname{MLP}
 \big((1+\gamma_m)\odot\operatorname{LN}_m(\widetilde H)+\beta_m\big).
\end{align}
Condition vectors are broadcast over tokens. Both layer normalizations have no learned affine parameters and use $\epsilon=10^{-6}$. The feed-forward MLP has dimensions $768\!\to\!3072\!\to\!768$ and tanh-approximated GELU activation. The modulation layer is zero-initialized, including its bias, giving zero initial residual gates~\citep{peebles2023dit}.

\paragraph{Long skips and output head.}
Let $H_\ell$ be the output of block $\ell$. Outputs $H_1,\ldots,H_4$ are cached. Before block $\ell\in\{5,6,7,8\}$, its current input $H_{\ell-1}$ is concatenated with $H_{9-\ell}$ and projected from 1536 to 768 dimensions using a separate learned linear layer. This implements skips $4\!\to\!5$, $3\!\to\!6$, $2\!\to\!7$ and $1\!\to\!8$~\citep{bao2023uvit}. In particular, block 5 receives the current and cached copies of $H_4$. The final token-wise layer normalization is modulated by a condition-dependent shift and scale, followed by a $768\!\to\!3$ output projection. Both the final modulation layer and output projection are zero-initialized. Training can use activation checkpointing within blocks.

\paragraph{Diffusion schedule and optimisation.}
Training uses a 1{,}000-step scaled-linear variance schedule with $\beta_t$ interpolated between $10^{-4}$ and $2\times10^{-2}$, an $\epsilon$-prediction objective, AdamW with learning rate and weight decay $10^{-4}$ and moment coefficients $(0.9,0.999)$, linear warmup over 1{,}000 updates, BF16 mixed precision, a micro-batch of four with eight gradient-accumulation steps (effective batch 32) and gradient-norm clipping at 1.0. Neither exponential moving averages nor early stopping were used. Unconditional pretraining lasts 60{,}000 updates. Conditional training starts from these weights with a reset optimizer, initially runs for 60{,}000 updates, and continues to 300{,}000 conditional updates with the optimizer state, data order, objective, learning rate and noise schedule preserved. The evaluated inference configuration is specified in Appendix~\ref{app:verifier_protocol}.

\paragraph{Periodic interpretation.}
A cyclic grid shift permutes tokens while leaving their pairwise wrapped displacements unchanged. Shared token-wise operations, global conditioning and site-aligned skip fusion preserve this symmetry, so the complete network is equivariant to integer cyclic shifts along any combination of the three grid axes. This guarantee is specific to translations on the discrete $8^3$ torus and does not imply rotation, continuous-translation or lattice-basis equivariance. This construction motivates the cyclic-shift diagnostics in Appendix~\ref{app:periodic_audit}; physical realization is assessed separately.

\paragraph{Numerical verification.}
Numerical tests verified the cyclic-shift identity $f(T_{\mathbf{s}}z_t,t,y)=T_{\mathbf{s}}f(z_t,t,y)$ along each axis and under simultaneous multi-axis shifts within relative and absolute tolerances of $2\times10^{-5}$. Additional tests confirmed the absence of absolute position parameters, directional wrap-around indexing of the bias tables, exact equality between the unconditional model and the zero-initialized conditional model at initialization, gradient flow into the PLD conditioner, and rejection of non-finite or out-of-domain targets.

\subsection{Learned Constructor architecture and evaluation}
\label{app:constructor_architecture}

The Constructor is the learned structure-realization stage that follows a Periodic U-DiT pore-field sample. It maps a decoded four-channel $32^3$ signed-distance field (SDF) to a topology and discrete building-block assignments, and expands the highest-scoring assignments into a compact candidate set for a single generated field.

\paragraph{Architecture.}
The Node Transformer reads the decoded field as 512 spatial tokens through a non-overlapping $4^3$ projection. Two learned class tokens carry the inorganic and organic node heads, and a topology embedding conditions the shared representation. Four pre-norm Transformer layers of width 256, eight heads, feed-forward width 1024 and dropout 0.1 feed independent linear classifiers (approximately 3.98M parameters).

Training the node heads on an equal mixture of raw and reconstructed fields keeps the training input close to the reconstructed field that the model sees at inference, and improves accuracy at both heads (Table~\ref{tab:constructor_node_head}). Validation uses reconstructed fields, the predicted topology and a soft node-pair compatibility penalty.

The Edge Transformer conditions on the topology, both node assignments and the four-channel SDF. It trains on raw fields with ground-truth topology and node labels, which grounds each linker prediction in a specific node pair instead of the field alone. For the single-head comparison in Table~\ref{tab:constructor_edge_head}, both rows share the same 13{,}691-example validation split and the same reconstructed fields: the conditioned row receives ground-truth topology and node labels, and the SDF-only control receives neither, so the gap between the two rows isolates the contribution of component conditioning. The edge vocabulary contains 646 training classes, together with an \texttt{\_\_unk\_\_} output for out-of-vocabulary labels that is excluded from generation.

\begin{table}[t]
\centering\small
\begin{tabular}{p{0.68\linewidth}rr}
\toprule
Node head and training input & Top-1 & Top-5 \\
\midrule
Inorganic node, raw SDF & 46.27\% & 74.60\% \\
Inorganic node, raw/reconstructed mixture & 47.97\% & 75.52\% \\
Organic node, raw SDF & 96.96\% & 98.13\% \\
Organic node, raw/reconstructed mixture & 97.04\% & 98.17\% \\
\bottomrule
\end{tabular}
\caption{\textbf{Node prediction and input-domain adaptation.} Both variants share one initialization, schedule and training budget, and are evaluated on reconstructed fields with the predicted topology and a soft node-pair compatibility penalty ($\lambda=0.5$). Both heads are scored on the full 13{,}691-structure split.}
\label{tab:constructor_node_head}
\end{table}

\begin{table}[t]
\centering\small
\begin{tabular}{p{0.68\linewidth}rr}
\toprule
Edge input condition & Top-1 & Top-5 \\
\midrule
SDF-only control & 36.08\% & 72.38\% \\
Topology/node-conditioned & 45.92\% & 83.32\% \\
\bottomrule
\end{tabular}
\caption{\textbf{Component-conditioned linker identification.} Both models train under matched schedules and are evaluated at their best validation checkpoints on the same 13{,}691 reconstructed fields. The conditioned model receives ground-truth topology and node labels. Values are single-head classification accuracies.}
\label{tab:constructor_edge_head}
\end{table}

At inference, the topology is fixed at its Top-1 prediction. The Node stage produces the five highest-scoring joint node pairs, and the Edge stage predicts five conditional edge candidates for each pair. For a node pair $n$ and linker $\ell$, the combined score is
\begin{equation}
 S(n,\ell)=\log P(n\mid\widehat x,\widehat\tau)
 +\log P(\ell\mid n,\widehat\tau,\widehat x).
\end{equation}
The assignments are globally ranked by $S$, with duplicates removed, yielding at most 25 alternatives. The w/o verifier and MEF variant takes the global first-ranked assignment. Its node pair can therefore differ from the Node head's first-ranked pair, because the conditional linker score also contributes to selection.

\paragraph{MOFFUSION-style component reference.}
Table~\ref{tab:moffusion_constructor} reports a MOFFUSION-style ResNet Constructor that we retrained on our expanded framework corpus with our 646-class edge vocabulary. This reference uses raw ground-truth SDFs and ground-truth topology on the 13{,}691-structure test split. The Transformer studies above use reconstructed SDFs on the validation split and separately specify their conditioning inputs.
\begin{table}[t]
\centering\small
\caption{\textbf{MOFFUSION-style Constructor retrained on our corpus.} Component classification on 13,691 test structures using raw SDFs. Building-block heads receive ground-truth topology. The edge head uses the reproduced 646-class vocabulary. Node 2 includes the empty-node class.}
\label{tab:moffusion_constructor}
\begin{tabular}{llrrr}
\toprule
Model & Head & Top-1 (\%) & Top-3 (\%) & Top-5 (\%)\\
\midrule
Topology (10 epochs) & Topology & 96.98 & 99.08 & 99.32\\
BB (10 epochs) & Node 1 & 28.12 & 50.98 & 60.95\\
 & Node 2 & 96.33 & 96.74 & 96.96\\
 & Edge & 26.89 & 51.40 & 63.09\\
BB (40 epochs) & Node 1 & 37.38 & 59.04 & 67.66\\
 & Node 2 & 96.69 & 97.14 & 97.52\\
 & Edge & 23.77 & 49.65 & 62.99\\
\bottomrule
\end{tabular}
\end{table}

\subsection{Multi-Edit-Former architecture and training}
\label{app:mef_architecture}
\paragraph{Task formulation.}
Given a source linker with $n\le3$ selected substitution sites, source states $s=(s_1,\ldots,s_n)$, source measured PLD $d_0$ and requested target $t$, Multi-Edit-Former (MEF) ranks \emph{complete} destination assignments $a=(a_1,\ldots,a_n)$ over the selected sites. The formal destination vocabulary is $\{\mathrm{H},\mathrm{F},\mathrm{Cl},\mathrm{Br},\mathrm{Me},\mathrm{OH},\mathrm{NH_2},\mathrm{CN},\mathrm{CF_3},\mathrm{Ph}\}$; ethyl, methoxy and formyl groups are source-only states that can be retained or replaced but not newly installed. The complete Cartesian assignment space contains at most $10^3$ formal assignments before deterministic single-site and pairwise constructability masks are applied, so MEF scores the enumerated assignments that satisfy these chemical constraints.

\paragraph{Invariant encoding and assignment score.}
Each site is embedded by its source state together with eight invariant scalars relative to the two linker anchors; six pairwise features describe inter-site distance and relative axial/radial geometry, and a numeric context vector carries the normalized source measured PLD, anchor separation, atom count and the mean and standard deviation of the selected sites' axial coordinates. A four-layer, eight-head Transformer ($d=256$, feed-forward dimension 1024, dropout $0.1$) encodes one context token and the site set, with pair geometry entering site-to-site attention through a learned symmetric bias that is shared across heads in the evaluated model. Because the encoder consumes distances and anchor-relative scalar projections, it is invariant to global translation and rotation; it is not an SE(3)-equivariant coordinate network. The target encoder represents the requested PLD and its difference from the source through two normalized scalars and eight Gaussian radial basis functions centred at the integer targets from 3 to 10~\AA. Every legal assignment receives the score
\begin{equation}
S_\theta(a\mid s,x,d_0,t)
=\sum_i u_\theta\!\left(h_i,E_s'(s_i),E_a(a_i),q\right)
-\lambda_{\mathrm{edit}}d_H(a,s),
\end{equation}
where $E_a$ is the destination-state embedding, $d_H$ is the Hamming edit count and $\lambda_{\mathrm{edit}}=0.05$. Exact score ties are resolved by fewer edits and then by a stable assignment key. An auxiliary PLD residual head has zero weight and does not contribute to action selection.

\paragraph{Training objective.}
Training tasks are constructed from measured PLD endpoints, and only measured legal destinations enter the denominator; unmeasured assignments are not treated as negatives. A measured destination is positive when it lies within $0.5$~\AA{} of the target, or, when no destination reaches that window, when it lies within $0.15$~\AA{} of the best measured target error; for source--target gaps above $3$~\AA{}, direction-improving positives are retained when available. MEF is trained with a multi-positive listwise ranking loss
\begin{equation}
\mathcal L_{\mathrm{rank}}
=-\frac{1}{B}\sum_{b=1}^{B}
\log
\frac{\sum_m C_{bm}P_{bm}\exp S_{bm}}
{\sum_m C_{bm}\exp S_{bm}},
\end{equation}
where $C$ marks measured legal candidates and $P$ marks positives. It was warm-started from a geometry-aware first-stage checkpoint and calibrated for two epochs over $82{,}334$ target-augmented tasks ($1{,}287$ steps per epoch), with $4{,}366{,}260$ parameters, AdamW~\citep{loshchilov2019adamw} (learning rate $3\times10^{-4}$, weight decay $10^{-4}$) and gradient clipping at $1.0$.

\paragraph{Lineage-disjoint data partitions.}
\label{app:mef_split}
The historical MEF benchmark groups related linker assets into lineages using base-asset identity and graph matching, then assigns complete lineages to training, development or held-out partitions. The three partitions contain 92, 12 and 11 lineages, respectively, and 475, 58 and 47 framework contexts. All pairwise intersections are zero for linker lineages, base-linker keys and base-XYZ identities. Training tasks draw exclusively from training contexts, with candidate menus restricted to measured, chemically legal destination states in the corresponding context.

\begin{center}
\small
\begin{tabular}{lrrrrrr}
\toprule
Feature & Train & Dev & Held-out & Train$\cap$Dev & Train$\cap$Held & Dev$\cap$Held\\
\midrule
Framework contexts & 475 & 58 & 47 & 0 & 0 & 0\\
Linker lineages & 92 & 12 & 11 & 0 & 0 & 0\\
Base-linker keys & 92 & 12 & 11 & 0 & 0 & 0\\
Base-XYZ identities & 92 & 12 & 11 & 0 & 0 & 0\\
Topologies & 147 & 35 & 32 & 32 & 26 & 12\\
Topology--node scaffolds & 336 & 52 & 43 & 35 & 19 & 10\\
\bottomrule
\end{tabular}
\end{center}

This partitioning evaluates transfer to new linker lineages under partially shared framework environments. Of the 32 held-out topologies and 43 topology--node scaffolds, 26 and 19 also occur in training. Here, scaffold denotes the MOF topology--node combination. The historical held-out partition has also supported subsequent diagnostics. This training-data audit is distinct from the paired generated-source editing experiment in Appendix~\ref{app:mef_control}.

\paragraph{Continuous target conditioning.}
The target encoder receives numerical PLD values through normalized scalars and fixed Gaussian bases centred at 3--10~\AA. These fixed centres provide a continuous expansion of the target and its difference from the source PLD. The chemical assignment menu is discrete, while conditioning varies continuously with the requested pore diameter. The historical evaluation uses integer targets on the 3--10~\AA\ grid, obtained as $t=\operatorname{round}(d_{\mathrm{raw}})$ in target construction. This grid standardizes candidate evaluation and hit-rate reporting; it does not restrict the encoder to categorical target labels.

\paragraph{Execution and scope.}
The selected assignment is materialized by a deterministic source-preserving editor that supports H$\rightarrow$FG, FG$\rightarrow$H and FG$_1\rightarrow$FG$_2$ transitions; anchor and retained-backbone displacements must stay below $10^{-3}$~\AA, and failed chemistry, clash or structural checks count as failed proposals and do not trigger substitution by a lower-ranked action. The edited linker is reassembled and geometrically evaluated as described in Appendix~\ref{app:verifier_protocol}. Assembly and Zeo++ measurement are outside the differentiable training objective. The model receives source-state tokens, linker and site geometry, source measured PLD and the requested target, but no categorical topology/node inputs or additional base-edge descriptors. Its geometric information is therefore local to the edited linker. It is trained on measured PLD, with continuous numerical target conditioning and a standardized integer evaluation grid from 3 to 10~\AA{}. The resulting physical pore response is assessed through reassembly and Zeo++.

\begin{table}[t]
\centering\small
\caption{\textbf{MEF architecture and training configuration.} The same trained editor is used throughout inference.}
\label{tab:mef_settings}
\begin{tabular}{ll}
\toprule
Component & Setting\\
\midrule
Selected sites & $1$--$3$\\
Formal destination states & H, F, Cl, Br, Me, OH, NH$_2$, CN, CF$_3$, Ph\\
Source-only states & Et, OMe, CHO\\
Formal-only menu & at most $10^3$, before constructability masking\\
Site / pair features & 8 (or 6) scalars + 16-centre RBF of anchor or inter-site distances\\
Encoder & 4 layers, 8 heads, $d=256$, FFN 1024, dropout $0.1$\\
Pair attention bias & symmetric, shared across heads in the evaluated model\\
Assignment score & target-conditioned site terms $-0.05\times$ edit count\\
Calibration & $82{,}334$ tasks, 2 epochs, $1{,}287$ steps per epoch\\
Parameters & $4{,}366{,}260$\\
Optimizer & AdamW, lr $3\times10^{-4}$, weight decay $10^{-4}$, clip $1.0$\\
Benchmark target grid & integer PLDs 3--10~\AA\\
Training endpoint & measured PLD\\
Preservation gate & $10^{-3}$~\AA{} for anchors and retained backbone\\
\bottomrule
\end{tabular}
\end{table}

\section{Verifier-guided generation and refinement}
\label{app:verifier_protocol}

\subsection{Direct geometric verification}
Learned scores prioritize candidate recipes and linker edits, while Zeo++ evaluates their realized pore apertures~\citep{willems2012zeo}. For an assembled framework $C$, the verifier returns $V(C)=\operatorname{PLD}_{\mathrm{Zeo++}}(C)$, and candidate selection minimizes $e(C;d^\star)=|V(C)-d^\star|$. All structure-selection and edit-refinement decisions use measured apertures.

This role differs from pre-assembly property prediction. LLM4MOF uses MOF2Zeo to rank topology--node--linker combinations before realization and relaxation, then uses Zeo++ to characterize the resulting structures~\citep{nam2026llm4mof}. CoFi-MOF applies its verifier after atomic frameworks are available, so direct geometric measurement can guide selection and assess each linker modification without training an additional surrogate.

\subsection{Candidate selection and local refinement}
Algorithm~\ref{alg:cofi} describes one generation attempt. The PLD-conditioned model, trained for 300,000 updates, samples one pore field with 200-step deterministic DDIM. The learned Constructor predicts one topology, retains five joint inorganic/organic node-pair assignments and predicts five conditional linkers for each pair. The resulting assignments are deduplicated and ranked by their combined model scores before PORMAKE assembly. This procedure realizes alternative chemical structures from the same generated field. Successfully assembled structures with finite positive measured PLD enter verification-based selection; assembly failures remain recorded.

Refinement starts from the measured candidate closest to the target whose linker supports the permitted edits. This editable source can differ from the best initial pore match, which remains available for final selection. Every successfully materialized and assembled edit is measured by Zeo++; failed proposals remain recorded. Editing is performed even when the source already satisfies the reporting tolerance, allowing improvement at finer resolution.

The final output is the closest measured pore match among the original and edited structures, with ties favoring an original candidate. If no valid edit is available, the best original structure is retained; if no original structure has a valid measurement, the attempt fails. Retaining the initial candidates makes the best measured aperture error non-increasing within the attempt. The selected output is not fed back into a second editing round.
\begin{algorithm}[t]
\caption{CoFi-MOF: verifier-guided generation and linker refinement}
\label{alg:cofi}
\begin{algorithmic}[1]
\Require Target PLD $d^\star$; learned Constructor; at most $M$ linker proposals ($M{=}1$ in Table~\ref{tab:main})
\State Draw attempt noise $\xi\sim\mathcal N(0,I)$; $\widehat z_e\gets\operatorname{DDIM}_\theta(\xi,d^\star)$
\State Decode one SDF $\widehat x=D(Q(\widehat z_e))$
\State $\widehat\tau\gets\operatorname{PredictTopology}(\widehat x)$
\State $\mathcal N\gets\operatorname{NodeTransformer}(\widehat x,\widehat\tau)$ \Comment{five ranked node pairs}
\State $\mathcal P\gets\varnothing$
\For{each node pair $n\in\mathcal N$}
 \State $\mathcal L_n\gets\operatorname{EdgeTransformer}(\widehat x,\widehat\tau,n)$ \Comment{five conditional linkers}
 \State Add $(\widehat\tau,n,\ell)$ for $\ell\in\mathcal L_n$ to $\mathcal P$
\EndFor
\State Rank and deduplicate $\mathcal P$ using node and conditional linker scores
\State $\mathcal C\gets\operatorname{AssembleAndVerify}(\mathcal P)$
\If{$\mathcal C=\varnothing$} \State \Return failure \EndIf
\State $\mathcal E\gets\varnothing$; $\mathcal S\gets\operatorname{EditableSources}(\mathcal C)$
\If{$\mathcal S\neq\varnothing$}
 \State $C_s\gets\arg\min_{C\in\mathcal S}e(C;d^\star)$
 \State $\mathcal A\gets\operatorname{MEF}(C_s,V(C_s),d^\star;M)$ \Comment{ranked edits; unchanged assignments excluded}
 \State $\mathcal E\gets\operatorname{AssembleAndVerify}(\operatorname{Apply}(C_s,\mathcal A))$
\EndIf
\State $\mathcal F\gets\mathcal C\cup\mathcal E$
\State $C_{\mathrm{sel}}\gets\arg\min_{C\in\mathcal F}e(C;d^\star)$ \Comment{ties favor the original structure}
\State \Return $(C_{\mathrm{sel}},V(C_{\mathrm{sel}}))$ \Comment{assembled structure and measured PLD}
\end{algorithmic}
\end{algorithm}

\subsection{Component ablations}
The variants in Table~\ref{tab:ablations} share the same generated fields and learned Constructor. CoFi-MOF w/o verifier and MEF directly outputs the globally highest-scoring assignment under the combined node and conditional linker log-probabilities, without aperture-based selection or editing. Zeo++ is applied afterward to evaluate the output, without changing the selected structure. CoFi-MOF w/o MEF uses direct geometric verification to select among the Constructor's assembled alternatives. CoFi-MOF (MEF top-1) applies the highest-ranked MEF edit and compares its reassembled structure with the retained candidate pool under the measured selection rule; this is the configuration evaluated in Table~\ref{tab:main}. CoFi-MOF (MEF top-10) permits up to ten MEF proposals per attempt and is reported as an ablation in Table~\ref{tab:ablations}. Each attempt produces at most one output.

The comparison isolates the successive roles of measured structural selection and local correction across these configurations. Appendix~\ref{app:mef_control} separately evaluates learned edit ranking against random legal modifications from shared sources, using both first proposals and verified best-of-ten edited products.

\subsection{Direct verification runtime}
\label{app:verifier_runtime}
In a serial CPU benchmark over 40 structures from the MOFDiff and MOFFlow-2 assembled-CIF evaluation package, Zeo++~0.3 required an average of 316~ms per structure. The test used a shared server with 256 CPU cores and 503~GB RAM, without GPU resources. This benchmark characterizes the measurement stage used throughout verification.

\section{Datasets and edit-response supervision}
\label{app:dataset}
\label{sec:editing_data}

\subsection{MOF-Edit-DB construction}
\label{app:dataset_scope}
\label{app:dataset_edits}
MOF-Edit-DB connects chemically specified linker modifications to the pore responses of their realized frameworks. We expand the PORMAKE edge library~\citep{pormake2021} with carboxylate-based linkers informed by reviews of Zr-MOFs~\citep{bai2016zrmofs} and titanium coordination polymers~\citep{du2024titanium}. This carboxylate-based extension contributes additional building blocks to PORMAKE's structural assembly workflow and is distributed with the MOF-Edit-DB resources. Functional-group attachment and removal further diversify these linkers, following the broader linker-design strategies explored in inverse design and local optimization~\citep{yao2021smvae,cleeton2025dream}. Each edit records its substitution sites, fragment identity and parent-linker mapping. The full base-edge catalog contains 844 entries, comprising 695 geometry-asset entries and 149 crosswalk aliases. The Constructor's supervised prediction vocabulary uses the 646 linker classes observed in its training split. The catalog, training vocabulary and chemically distinct linker count describe different populations. Table~\ref{tab:dataset} separately summarizes base-edge and edited-edge resources.

A paired example associates a framework context $c$ and linker edit $a$ with
\begin{equation}
(c,a)\longmapsto(d_{\mathrm{parent}},d_{\mathrm{edited}},\Delta d),
\qquad \Delta d=d_{\mathrm{edited}}-d_{\mathrm{parent}}.
\end{equation}
The context identifies the topology, inorganic and optional organic nodes, and parent linker. Parent and edited frameworks share this chemical and topological context, while their realized geometries may differ. The resulting action--response representation supports learning how local chemical changes affect framework-level pore apertures.

\subsection{Pore labels, paired records and quality screening}
\label{app:dataset_protocol}
\label{app:dataset_pairs}
Frameworks are assembled with PORMAKE, relaxed with UFF4MOF in LAMMPS~\citep{addicoat2014uff4mof,thompson2022lammps}, and characterized with Zeo++~\citep{willems2012zeo}. The released pore labels therefore describe force-field-relaxed structures. PLD and LCD correspond to the Zeo++ $D_f$ and $D_i$ fields; PLD does not exceed LCD in any released record.

The master table contains 138,771 records: 26,058 parents and 112,713 edited frameworks. Among the edited records, 108,145 have a canonical parent PLD and a paired response; the remaining 4,568 retain their own pore labels but are excluded from paired-response statistics. Counts refer to records rather than independently deduplicated crystal structures. The collection spans 584 topology labels and 37,428 canonical scaffold keys. Table~\ref{tab:dataset} summarizes its framework and linker resources.
\begin{table}[t]
\centering
\small
\caption{MOF-Edit-DB v1.2. Framework counts are records; linker counts are catalog entries. Canonical paired responses are the subset used in Figure~\ref{fig:dataset}.}
\label{tab:dataset}
\begin{tabular}{lrp{0.48\linewidth}}
\toprule
Resource & Count & Role \\
\midrule
Framework records & 138,771 & Pore labels with framework context \\
\quad Parent records & 26,058 & Reference structures \\
\quad Edited records & 112,713 & Products of annotated linker modifications \\
Canonical paired responses & 108,145 & Before/after PLD supervision \\
Edited-edge catalog & 78,961 & Edit identities and asset mappings \\
Base-edge catalog & 844 & 695 asset entries and 149 aliases \\
\bottomrule
\end{tabular}
\end{table}

Response statistics use the master table's \texttt{delta\_pld}, which agrees with \texttt{modified\_pld - base\_pld} within $10^{-4}$~\AA\ wherever populated. Canonical identifiers and scaffold mappings define the parent--edit associations; auxiliary comparisons with alternative parent references are documented separately in the data README.

\label{app:dataset_qc}
MOFChecker-based screening addresses atomic overlaps and detached molecular fragments~\citep{jin2025mofchecker}. The released quality annotations contain no positive overlap or lone-molecule flags, although these individual flags are unrecorded for 201 and 1,731 records, respectively. The retention policy includes 9,729 overcoordination-only records and 342 records flagged as not three-dimensionally connected. These annotations support stricter task-specific subsets without treating every retained record as satisfying an identical set of structural checks.

\subsection{Pore coverage and edit responses}
\label{app:dataset_stats}
\begin{figure}[t]
\centering
\includegraphics[width=\linewidth]{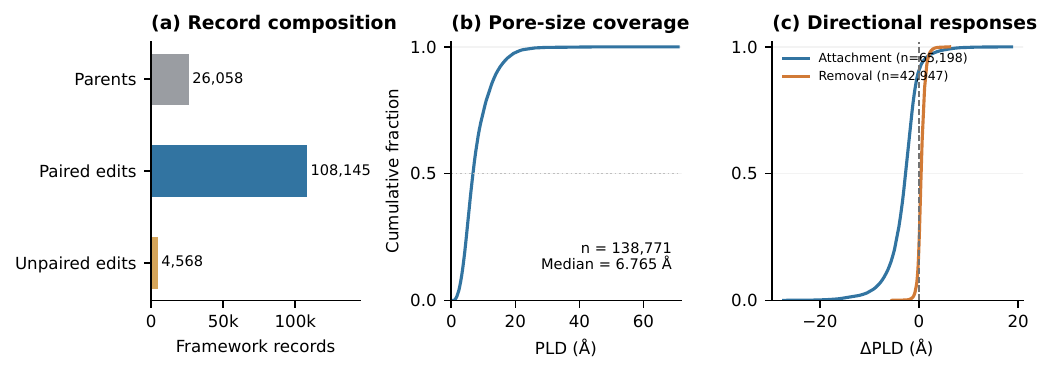}
\caption{\textbf{Coverage and edit responses in MOF-Edit-DB.} (a) Parent records, paired edits and edits without a canonical parent comparison. (b) PLD coverage over all 138,771 records. (c) Canonical parent-to-edited $\Delta$PLD for functional-group attachment and removal, normalized separately. Curves show empirical cumulative distributions over the full observed ranges; the vertical line denotes no change. All response statistics use the master table.}
\label{fig:dataset}
\end{figure}

Using each record's canonical \texttt{modified\_pld}, 72,855 of 138,771 records (52.5\%) have $0<\mathrm{PLD}<7$~\AA. All records have finite positive values, so this proportion also describes the full positive-PLD population. The collection therefore provides substantial coverage of the small-aperture regime together with chemically specified edit responses. Across all records, PLD has mean 7.933~\AA\ and median 6.765~\AA, with 5th and 95th percentiles of 2.548 and 17.104~\AA. Figure~\ref{fig:dataset} shows the full empirical distributions. The canonical pairs comprise 65,198 functional-group attachment examples and 42,947 removal examples. Attachment has median $\Delta$PLD $-2.709$~\AA, with 9.1\% positive responses; removal has median $+0.423$~\AA, with 77.8\% positive responses.

These contrasting shifts provide examples of both pore contraction and expansion, while the within-direction variation makes the framework context informative: the edit category alone does not determine the sign or magnitude of the response. The two edit groups contain different parent populations, so their distributions describe the sampled actions rather than a matched comparison of attachment and removal.

\subsection{Learning tasks and data access}
\label{app:dataset_usage}
The paired records support edit-response prediction, target-conditioned action ranking, and transfer evaluation across scaffold or linker families. Family-disjoint evaluation should keep each parent and its related edits together; a standardized public benchmark split is not supplied in the current release. The review package provides data tables, edit catalogs, quality annotations and a structure manifest, with geometry availability documented in the accompanying README. The compact package does not contain every referenced geometry.

MOF-Edit-DB is distinct from the global generator's mixed training corpus, which combines selected small-pore MOFFUSION structures~\citep{park2025moffusion} with self-generated frameworks. Its release size is not the training-set size of each CoFi-MOF component. MEF is trained on PLD labels measured from assembled structures (Appendix~\ref{app:mef_architecture}).

\subsection{Data underlying the PLD motivation figure}
\label{app:pld_motivation}
Figure~\ref{fig:pld_motivation} compares the selected experimental CoRE collection with the processed MOFFUSION collection and a BW-DB release derived from the hypothetical MOFs of \citet{boyd2019bwdb}. For CoRE MOF 2019 v2.0.1, we use the source tables \texttt{ASR\_20241119.csv} and \texttt{FSR\_20241119.csv}, retaining entries marked \texttt{CR=Y}. The 6,603 ASR and 3,764 FSR entries have finite positive PLDs. Pooling them gives 10,367 entries, of which 9,173 (88.5\%) lie strictly below 7~\AA. ASR and FSR denote removal of all solvents and free solvents, respectively~\citep{chung2019core}; the pool counts preparation variants rather than unique experimental frameworks. The processed MOFFUSION table contains 247,744 rows, with 247,732 finite positive PLDs and 12 missing or invalid values; 53,788 positive values (21.7\%) are below 7~\AA.

For BW-DB, we analyze \texttt{MetalOxo.lmdb} from the \texttt{bw\_db.tar.gz} release distributed with MOFDiff\footnote{\url{https://zenodo.org/records/10806179}}. This release contains 308,962 records, of which 308,961 have a PLD value. Excluding the 39 nonpositive values leaves 308,922 positive-PLD entries. Among these, 72,797 satisfy $0<\mathrm{PLD}<7$~\AA, giving 23.56\%. BW-DB is an upstream structural resource for MOFDiff, MOFFlow and MOFFlow-2~\citep{fu2024mofdiff,kim2025mofflow,kim2025mofflow2}. These methods apply their own preprocessing and data selection, so the figure characterizes the analyzed BW-DB collection rather than each model's final training split.

All percentages use the corresponding positive-PLD population and are calculated from counts below the strict 7~\AA\ threshold, independently of curve smoothing. The curves use a reflected Gaussian kernel with bandwidth 0.25~\AA\ and a 0.02~\AA\ evaluation grid. Normalization retains the full positive-PLD population, including the tail beyond the displayed 20~\AA\ range. Source-reported descriptors and preparation settings are retained.

The contrast is consistent across both hypothetical collections, strengthening the motivation for dedicated generation in the experimentally prevalent small-aperture regime. These statistics describe aperture coverage through PLD, rather than the size of every cavity within a framework.

\subsection{\revised{Sources and construction of the global generation corpus}}
\label{app:global_corpus}

The global generation corpus contains 136,911 records (109,529 for training): 65.8\% from the MOFFUSION small-pore subset and 34.2\% newly constructed (Table~\ref{tab:global_corpus_sources}). It is distinct from the 138,771-record MOF-Edit-DB.

\begin{table}[t]
\centering\small
\caption{\revised{\textbf{Sources of the global generation corpus.} Percentages are rounded independently.}}
\label{tab:global_corpus_sources}
\begin{tabular}{lrr}
\toprule
Source & Records & Share (\%)\\
\midrule
MOFFUSION-derived small-pore subset & 90,130 & 65.8\\
New construction, round 1 & 9,774 & 7.1\\
New construction, round 2 & 37,007 & 27.0\\
\bottomrule
\end{tabular}
\end{table}

New structures combine topology and node types from the original resource with 476 additional carboxylate base linkers included in MOF-Edit-DB, subject to connectivity and recipe checks. The two rounds prescreen 200,000 and 500,000 candidate recipes, respectively. A PLD proxy selects 50,000 candidates per round using approximately balanced bins over predicted PLD ranges of $[0,12)$ and $[0,9)$~\AA. Candidates then undergo PORMAKE assembly, LAMMPS/UFF4MOF relaxation and Zeo++ measurement. Retained new records have measured PLD in $[0,15)$~\AA{} and corresponding four-channel SDFs. The proxy is used for prescreening; the final pore labels are measured values.

\section{Experimental setup and evaluation}
\label{app:evaluation}
\subsection{Structural realization and pore measurement}
\label{app:physical_eval_protocol}
Recipes are assembled with PORMAKE~\citep{pormake2021} using topology, inorganic and optional organic node, and linker identifiers. Augmented-edge geometries supply building blocks absent from the original library. Frameworks are written as P1 CIFs with assembly-defined bonds. A minimum-image check excludes structures with any atom pair closer than 0.1~\AA\ before pore characterization.

\revised{We use PORMAKE v0.2.3. Aligning a two-connected linker's connection points fixes its axis but leaves rotation about that axis. The PORMAKE documentation describes random axial rotations and a fork addressing them.\footnote{\url{https://github.com/Sangwon91/PORMAKE\#new-fork}} Such orientations can affect pore-window geometry. Zeo++ measures each successfully assembled, measurable candidate from its atomic coordinates, so its realized orientation and cell adjustment enter the PLD used for selection.}

Zeo++ provides the largest cavity diameter, pore-limiting diameter and largest free-path diameter. Finite positive PLD determines eligibility for target-error selection; missing and non-finite measurements are failures rather than zero-valued PLDs.

Supplementary relaxation uses UFF4MOF generated by \texttt{lammps-interface}, with \texttt{units real}, \texttt{atom\_style full} and periodic boundaries. Three conjugate-gradient stages perform fixed-cell minimization, isotropic cell relaxation to zero pressure, and final fixed-cell minimization. Each uses energy and force tolerances of $10^{-6}$ and $10^{-8}$, with a 180-second wall-clock limit. This protocol is used only in the downstream screening analysis of Appendix~\ref{app:relaxed_realizations}.

\subsection{Evaluation populations and statistics}
\label{app:evaluation_metrics}
Targets are $3,4,5,6,7$~\AA, with 1,000 allocated generation attempts per target. Hit rates use all allocated attempts as the denominator; assembly and measurement failures count as misses. Hit@$\,\tau$ denotes an absolute PLD error no greater than $\tau$; MAE and signed bias are calculated over selected structures with finite positive PLD.

The initial batch contains 100 attempts per target, followed by a continuation allocating 900 per target. Pooled rates aggregate counts across these batches. Paired uncertainty for the component ablations and editing comparisons resamples shared noise identifiers using a block bootstrap with 2,000 resamples and a fixed seed; batch-specific intervals are reported separately. The periodic-backbone comparison uses its own 500-pair design and uncertainty calculation, described in Appendix~\ref{app:periodic_audit}.

Within-batch structural uniqueness measures distinct frameworks among measured outputs and is reported only for the 500-attempt batch evaluated by structure matching. It is distinct from novelty relative to a training collection. Baseline-specific sampling, selection and deduplication are detailed below; the table populations retain those distinctions.

\subsection{Baseline configurations}
\label{app:baseline_settings}
The external-model comparison evaluates targets of 3--7~\AA\ with 1,000 allocated attempts per method and target. For MOFFlow-2 and MOFDiff, pore characterization uses the retained assembled CIFs from the formal runs: 911--938 and 899--932 assembled structures per target, respectively, with finite positive PLDs. MOFFUSION uses the retained raw assemblies from its target-conditioned validation runs; six target-3 structures have zero PLD and are excluded from the finite-positive population.

\paragraph{GA: target-conditioned search.}
The genetic algorithm searches combinations of 30 topologies, 208 metal-cluster nodes and 3,334 linkers, using reference data curated for the 0--7~\AA\ pore range. Its target-dependent fitness is
\begin{equation}
 \mathrm{fitness}(g,d^\star)=\exp\!\big(-((\mathrm{PLD}(g)-d^\star)/0.5)^2\big),
\end{equation}
where PLD is obtained from a measured library or an on-demand Zeo++ evaluation. The population size is 24, with four elites, tournament size three, single-field crossover probability 0.8 and single-field mutation probability 0.25. Each target is allocated 1,000 scored proposals.

Table~\ref{tab:main} evaluates the full combinatorial search space using 1,000 scored proposals per target. Repeated proposals are retained in all attempt-level metrics, with cached geometric measurements reused for repeated combinations. Failed proposals count as misses, and MAE and bias are averaged over proposals with finite positive measured PLD, retaining their multiplicities. Across targets of 3--7~\AA, the 1,000 proposals contain 253, 222, 236, 213 and 232 distinct topology--node--linker combinations, respectively, corresponding to repetition rates of 74.7--78.7\%. Of these distinct combinations, 144, 154, 146, 152 and 149 have finite positive PLD. The run required 883 new geometric measurements, and 99.9\% of its unique combinations were absent from the preloaded measured library. These combination counts characterize repetition separately from the attempt-level targeting results.

\paragraph{MOFDiff: target-directed pool selection.}
We adapt MOFDiff~\citep{fu2024mofdiff} using its building-block diffusion model and a separately trained PLD prediction head. For each requested target, the head scores 15,351 posterior latents from the official held-out validation split, and the 1,000 latents with smallest predicted target error are decoded and assembled. Zeo++ characterizes these assemblies from their atomic coordinates.

This adaptation is labeled \emph{MOFDiff + selection}: its target-directed retrieval-and-realization procedure uses a PLD predictor to choose promising latent candidates before decoding and assembly. Thus, the comparison includes predictor-guided candidate selection rather than unselected unconditional samples. Target subsets may overlap; targets 3 and 4 share 780 selected latents. Each selected latent counts as one final-stage attempt, with shared-pool scoring constituting an additional inference cost.

\paragraph{MOFFlow-2: learned PLD conditioning.}
The MOFFlow-2 adaptation~\citep{kim2025mofflow2} uses the released data and checkpoints, initializing the sequence generator from its unconditional checkpoint and fine-tuning continuous PLD conditioning through property embeddings and cross-attention. Labels use the database's \texttt{largest\_free\_sphere\_diameter [A]} field; the filtered training and validation sets contain 187,028 and 10,242 structures. Sequence-model fine-tuning lasts two epochs, using checkpoint \texttt{epoch\_1-step\_17540-loss\_0.1640.ckpt}, while the structural generator retains its released weights.

Each target produces 1,000 attempts in five batches of 200. Generation and measurement failures remain in the allocated denominator. This adaptation uses a trained condition rather than posterior-pool retrieval or target-based output selection.

\paragraph{MOFFUSION: conditional pore-field generation.}
The MOFFUSION adaptation~\citep{park2025moffusion} is trained on the mixed corpus for 300,000 cumulative updates. Its target condition is scaled by $1/100$. Sampling uses 200-step deterministic DDIM with seed $20260915+\text{target}$. For the fixed-upstream comparison, the original top-5 expansion is increased to 25 ranked building-block assignments under the predicted topology. Candidate recipes are assembled with PORMAKE and evaluated with Zeo++; the realized structure with PLD closest to the requested target is retained. The baseline therefore includes measured candidate selection rather than rank-1 realization alone. The experiment generates 1,000 attempts at each integer target from 3 to 11~\AA; Table~\ref{tab:main} evaluates targets 3--7~\AA; Appendix~\ref{app:budget25} reports the candidate-level accounting for the 25-realization expansion.

\subsection{\revised{Cross-corpus compatibility and comparison scope}}
\label{app:training_compatibility}

MOFDiff and MOFFlow-2 retain their released training distributions and native structural realizers in Table~\ref{tab:main}; full common-corpus retraining is not established by this comparison. Their native representations also impose building-block-count, atom-vocabulary and metal-template constraints, so format conversion alone does not preserve the CoFi-MOF corpus's chemical coverage.

We therefore examined partial adaptation on compatible subsets. MOFDiff's building-block cap was expanded to 64 while retaining its released building-block encoder and library, with adaptation of its coarse-grained model and PLD head. MOFFlow-2 adapted only its PLD-conditioned sequence generator, retaining the released three-dimensional flow and metal-template library. These partial updates did not yield effective transferred baselines under the native realization pipelines, with fixed-upstream best-of-25 Hit@0.5 of 12.00\% for MOFDiff and 8.00\% for MOFFlow-2 over 40 groups per target. These measurements diagnose partial transfer, rather than full common-corpus retraining or the limits of broader architectural adaptation.

Accordingly, we interpret the external-model rows as complete-pipeline comparisons under their respective training distributions. The same-corpus Periodic U-DiT versus 3D U-Net study in Appendix~\ref{app:periodic_audit} provides the controlled comparison of the deployed backbone contribution.

\subsection{\revised{Implementation of the 25-candidate verifier budget}}
\label{app:budget25}

The four neural pipelines in Table~\ref{tab:main} share a fixed-upstream realization budget. We fix 1{,}000 upstream
outputs per target before any measurement, expand each output into 25 candidate structures,
measure every candidate once with Zeo++ (\texttt{network -ha -res}) on the assembled
structure, and retain the candidate with the smallest measured $|\mathrm{PLD}-t|$. Upstream lists
were fixed before evaluation. No attempt was replaced, resampled or retried, and failed evaluations remain in
the denominator and count as misses. The methods differ in which degrees of freedom the 25 candidates
explore, so this is not an identical search space, training-compute or runtime comparison. GA uses its separately specified 1{,}000-proposal search protocol.

\paragraph{MOFFlow-2.} One upstream output is a fixed building-block sequence. The 25 candidates are
25 independent three-dimensional realizations of that sequence obtained through the official
inference interface. Randomness enters through the prior over building-block translation, random
rotations, rotatable-bond torsion angles in $[-\pi,\pi]$, and log-normal cell lengths with cell angles
in $[60^\circ,120^\circ]$ followed by 50 integration steps. Building-block identity is fixed, so the
budget explores continuous geometry only. The 25 realized CIFs of an upstream output were verified to
be distinct.

\paragraph{MOFDiff.} One upstream output is a fixed coarse-grained prediction together with its
KD-tree building-block assignment. The 25 candidates are 25 assembly initializations of that same
prediction. We exposed the annealing seed, which the released script fixes at zero, and used seeds
$0,\dots,24$, so that only the initial building-block orientations and the three-round annealing
trajectory change.

\paragraph{MOFFUSION.} One upstream output is a generated pore field. The topology classifier returns
its top-1 prediction, and the building-block predictor is expanded to the 25 highest joint
log-probability node and edge assignments under that topology, deduplicated. Each assignment is
assembled with PORMAKE and measured. The budget therefore explores discrete building-block assignments
under a fixed topology, the closest available analogue to our Constructor. The top-25 expansion is a
budget added for this comparison rather than a released default.

\paragraph{CoFi-MOF.} The initial pool contains the Constructor's 25 ranked building-block assignments for
one generated pore field. MEF then edits the selected source structure; the MEF top-1 arm adds its
highest-ranked edit, which competes with the 25 verified candidates under the same selection
rule. Thus, the full CoFi-MOF row uses 25 fixed-upstream realizations plus one measured edit. The MEF top-10 setting is an ablation (Table~\ref{tab:ablations}): it spends up to nine additional measured candidates per attempt and is therefore excluded from the fixed-upstream comparison.

\paragraph{Recorded call accounting.} Each neural pipeline has a nominal allocation of 125{,}000 fixed-upstream realizations; CoFi-MOF additionally allocates one MEF edit per attempt. Within
each method, recorded assembly and Zeo++ calls are 105{,}776 assembled candidates and 105{,}711 finite
PLDs for MOFFUSION, 124{,}850 assembly calls and 124{,}211 Zeo++ calls for the CoFi-MOF verifier, and
115{,}510 and 114{,}374 finite PLDs for the MOFDiff and MOFFlow-2 candidate sets.
These counts describe each pipeline and are not additive across methods, because candidate pools are
shared within a run and the assembly and measurement environments differ.

\section{Additional results}
\label{app:additional_results}
The following analyses examine periodic consistency and structural realization, the roles of geometric verification and local editing, and the resulting pore-targeting accuracy.

\subsection{Periodic consistency and structural realization}
\label{app:periodic_audit}
We examine whether explicitly periodic latent generation improves downstream MOF realization. We compare the MOFFUSION 3D U-Net against unconditional Periodic U-DiT (Appendix~\ref{app:periodic_architecture}). Figures denote these models by B0 and B1, respectively, and the PLD-conditioned variant by D2. The comparison controls the data, latent representation, objective, optimizer and training budget: both backbones were trained for 60{,}000 updates on the same 109{,}529 training structures encoded by the same frozen VQ-VAE encoder, using the $\epsilon$-prediction objective, AdamW ($10^{-4}$ learning rate and weight decay, 1{,}000 warmup updates, gradient-norm clip 1.0) and an effective batch size of 32 in BF16 on one RTX 4090, under one common seed. The test split was not accessed.

\paragraph{Matched sampling and paired statistics.}
Evaluation draws on a bank of 500 shared Gaussian noise tensors: attempt $i$ from each backbone start from the same initial noise and are decoded with deterministic DDIM (200 steps, $\eta=0$), then pass through an identical assembly--relaxation--Zeo++ chain with no stage-specific changes. Paired design removes variation from the diffusion noise draw, so stage-wise differences are assessed with the exact McNemar test on discordant pairs, and the final-validity difference is summarized with a paired-bootstrap 95\% confidence interval over 10{,}000 replicates.

\begin{figure}[t]
\centering
\includegraphics[width=\linewidth]{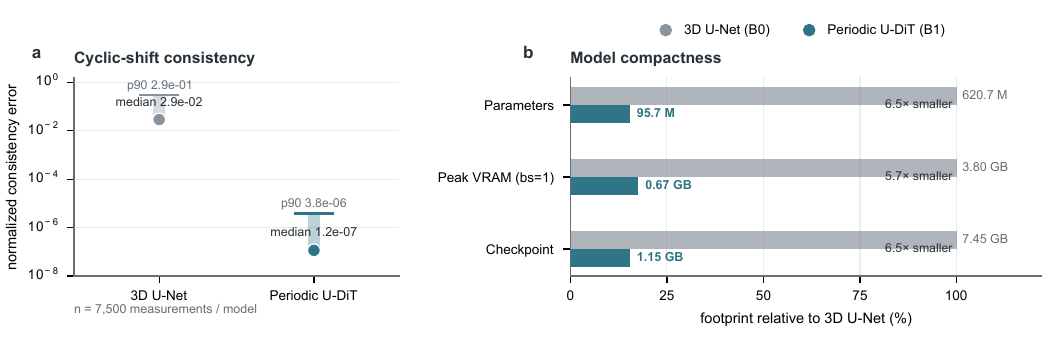}
\caption{\textbf{Denoiser periodicity and model footprint.} (a) Normalized cyclic-shift consistency error over 7{,}500 probes per model (medians and 90th percentiles). (b) Parameter count, peak forward memory (batch size 1) and checkpoint size relative to the 3D U-Net.}
\label{fig:periodic_shift}
\end{figure}

\paragraph{Cyclic-shift and uniform-input probes.}
For shift $T_\Delta$, we measure
\begin{equation}
 E_{\mathrm{eq}}=
 \frac{\|\epsilon_\theta(T_\Delta z_t,t,c)-T_\Delta\epsilon_\theta(z_t,t,c)\|_2}
 {\|\epsilon_\theta(z_t,t,c)\|_2+10^{-12}}.
\end{equation}
The protocol uses 500 held-out validation latents, timesteps $\{50,500,950\}$ and five shifts, yielding 7{,}500 measurements per model. The unconditional and PLD-conditioned periodic models have median relative errors near $10^{-7}$ (Table~\ref{tab:periodic_probes} and Figure~\ref{fig:periodic_shift}), consistent with cyclic-shift equivariance up to numerical precision. A complementary uniform-input probe broadcasts the per-channel validation-latent mean to all sites; at $t=50$ the coefficient of variation of token-wise output norms is 0.4524 for the 3D U-Net and below $10^{-5}$ for both periodic models.

\begin{table}[t]
\centering\small
\caption{\textbf{Spatial consistency of denoiser outputs.} Relative cyclic-shift error is aggregated over 7,500 probes. Uniform-input CV is measured at $t=50$. }
\label{tab:periodic_probes}
\begin{tabular}{lccc}
\toprule
Model & Mean $E_{\mathrm{eq}}$ & Median $E_{\mathrm{eq}}$ & Uniform-input CV\\
\midrule
3D U-Net & $1.034\times10^{-1}$ & $2.859\times10^{-2}$ & $4.524\times10^{-1}$\\
Periodic U-DiT & $1.227\times10^{-6}$ & $1.156\times10^{-7}$ & $6.105\times10^{-6}$\\
PLD-conditioned U-DiT & $5.024\times10^{-7}$ & $1.067\times10^{-7}$ & $2.811\times10^{-6}$\\
\bottomrule
\end{tabular}
\end{table}

\begin{figure}[t]
\centering
\includegraphics[width=\linewidth]{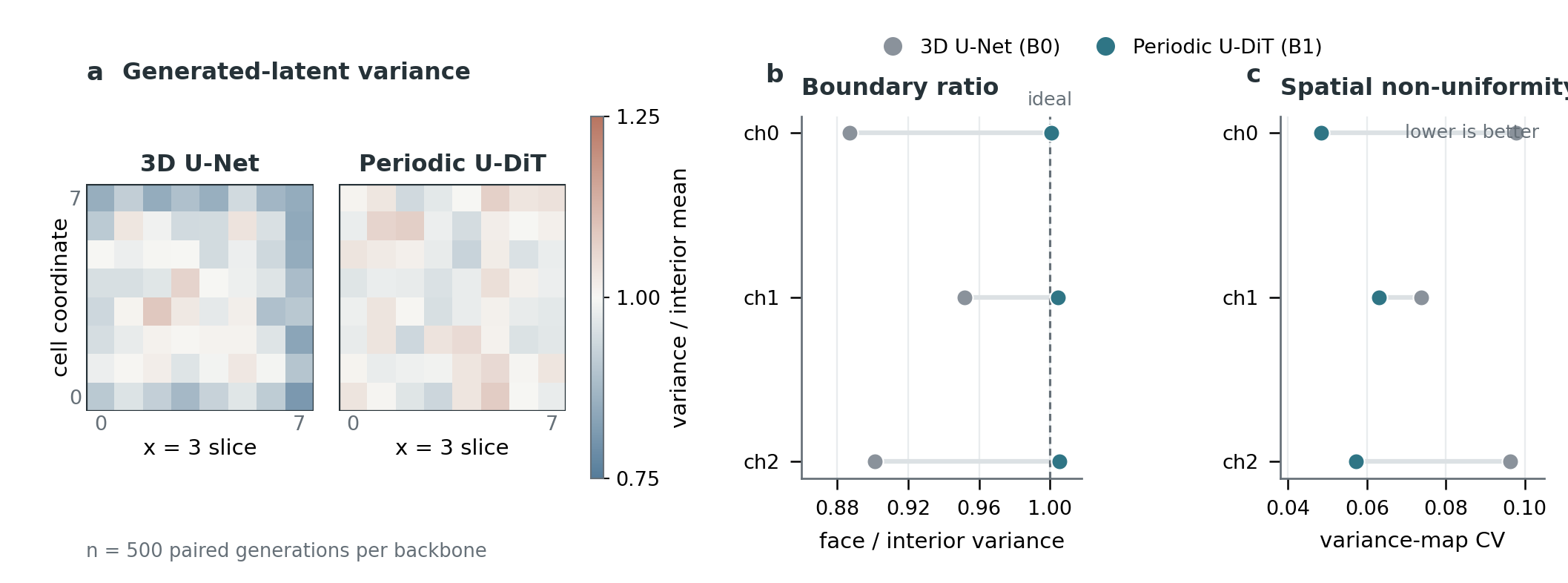}
\caption{\textbf{Generated-latent spatial statistics over 500 paired generations.} (a) Per-site variance maps ($x{=}3$ slice). (b) Face-to-interior variance ratio per channel; cyclic-shift-invariant sampling gives a population ratio of 1, with finite-sample deviations. (c) Variance-map coefficient of variation; lower is more spatially uniform.}
\label{fig:periodic_variance}
\end{figure}

\paragraph{Generated latent fields.}
Across the 500 paired generations, per-site variance over samples reveals whether boundary artefacts propagate into the generated distribution. The 3D U-Net suppresses variance near the unit-cell faces, with face-to-interior ratios $(0.8870,0.9518,0.9014)$ across the three latent channels, whereas Periodic U-DiT closely matches faces and interior at $(1.0008,1.0043,1.0053)$ and reduces the mean variance-map coefficient of variation from 0.0893 to 0.0561 (Figure~\ref{fig:periodic_variance}). The sampled latent statistics thus complement the equivariance probes, showing substantially more uniform spatial variation across the unit cell.

\begin{figure}[t]
\centering
\includegraphics[width=\linewidth]{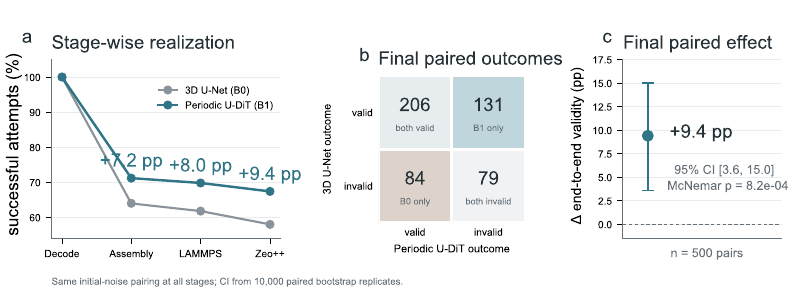}
\caption{\textbf{Paired structural realization on 500 shared initial noises.} (a) Stage-wise success. (b) Discordant final outcomes; discordant pairs determine the McNemar test. (c) End-to-end validity difference with the paired-bootstrap 95\% CI.}
\label{fig:paired_realization}
\end{figure}

\paragraph{Paired structural realization.}
Both models decode all 500 latents successfully, so the comparison emerges only after realization. Successful attempts increase from 320 for the 3D U-Net to 356 for Periodic U-DiT at assembly, from 309 to 349 after LAMMPS relaxation, and from 290 to 337 after Zeo++ characterization (Figure~\ref{fig:paired_realization}). End-to-end validity rises from 0.580 to 0.674 ($+9.4$ percentage points; paired-bootstrap 95\% CI $[3.6,15.0]$; one-sided exact McNemar $p=8.2\times10^{-4}$). The difference is already evident at assembly and persists through the subsequent stages. Discordant final outcomes include 84 successes unique to the 3D U-Net and 131 unique to Periodic U-DiT, yielding 47 additional valid structures.

\paragraph{Resource footprint.}
Periodic U-DiT is also substantially more compact: 95.7 M parameters versus 620.7 M for the 3D U-Net ($6.5\times$ smaller), 0.67\,GB peak forward memory versus 3.80\,GB ($5.7\times$), and a 1.15\,GB checkpoint versus 7.45\,GB (Figure~\ref{fig:periodic_shift}b). A batch-one forward pass of Periodic U-DiT uses 668.5\,MB peak memory and 69.40\,GFLOPs.

\paragraph{Interpretation.}
Together, the probes establish periodic spatial consistency, and the paired comparison shows improved structural realization for the complete architecture. Backbone family, width and skip structure change alongside the attention bias, so the comparison supports the architecture as deployed. The results cover one training seed and one matched-noise panel. At 60,000 updates, denoising MSE is 0.07796 for the 3D U-Net and 0.08062 for Periodic U-DiT, so the gain comes from realization quality at a fixed training budget rather than from a lower training loss.

\subsection{Geometric verification and local pore refinement}
\label{app:rigid_results}
CoFi-MOF w/o verifier and MEF, recorded as the direct output in the archived run summaries, retains the learned Constructor and returns the global top-ranked assignment under the joint node--linker score. CoFi-MOF w/o MEF retains geometric verification but omits local linker refinement.
\paragraph{Evaluation coverage.}
The initial 500-attempt batch and the subsequent 4,500-attempt continuation cover targets of 3--7~\AA, for 5{,}000 allocated attempts. Table~\ref{tab:ablations} uses all allocated attempts as hit-rate denominators, with assembly and measurement failures counted as misses. Metric definitions and pooling are given in Appendix~\ref{app:evaluation_metrics}. The target-wise results below describe the MEF top-10 ablation; the MEF top-1 configuration evaluated in Table~\ref{tab:main} is discussed in Section~\ref{sec:ablations}.

\paragraph{Target-wise improvement.}
In the MEF top-10 ablation, MEF increases Hit@0.5 at every target, adding 71, 50, 38, 33, and 45 hits at 3--7~\AA, respectively. The final rates range from 94.50\% at 3~\AA\ to 77.86\% at 7~\AA. The pooled gain is 237 hits (4.75 percentage points); at 0.2~\AA, it is 461 hits (9.23 points). The MEF top-1 configuration of Table~\ref{tab:main} keeps the same coarse pool and applies one MEF edit, adding 143 hits at 0.5~\AA{} and 170 at 0.2~\AA{} (Section~\ref{sec:ablations}). Because the coarse pool is retained under both configurations, no coarse-pool hit is displaced. Paired 95\% intervals for the MEF top-10 Hit@0.5 gain are [4.20, 8.80] percentage points in the initial batch and [3.92, 5.28] in the continuation, using shared-noise blocks.

\paragraph{Editing coverage.}
Across the 5,000 attempt manifests, 4,713 identify an editable source, 281 report no editable source, and six have no measured coarse candidate. These are source-availability categories, not selected-output success counts. An editable source is selected from the Constructor's measured structures. When no eligible linker modification is available, the model retains the best original pore match.

\paragraph{Structural uniqueness.}
In the initial 500-attempt batch, structural uniqueness is 436/476 (91.6\%) for CoFi-MOF w/o verifier and MEF, 490/499 (98.2\%) for CoFi-MOF w/o MEF, and 493/499 (98.8\%) for CoFi-MOF with MEF. These measurements describe distinct structures among measured outputs in that batch, not novelty relative to a training collection. The uniqueness analysis is restricted to this initial batch.

\subsection{Representative linker edits}
\label{app:edit_cases}
\begin{figure}[t]
\centering
\includegraphics[width=0.9\linewidth]{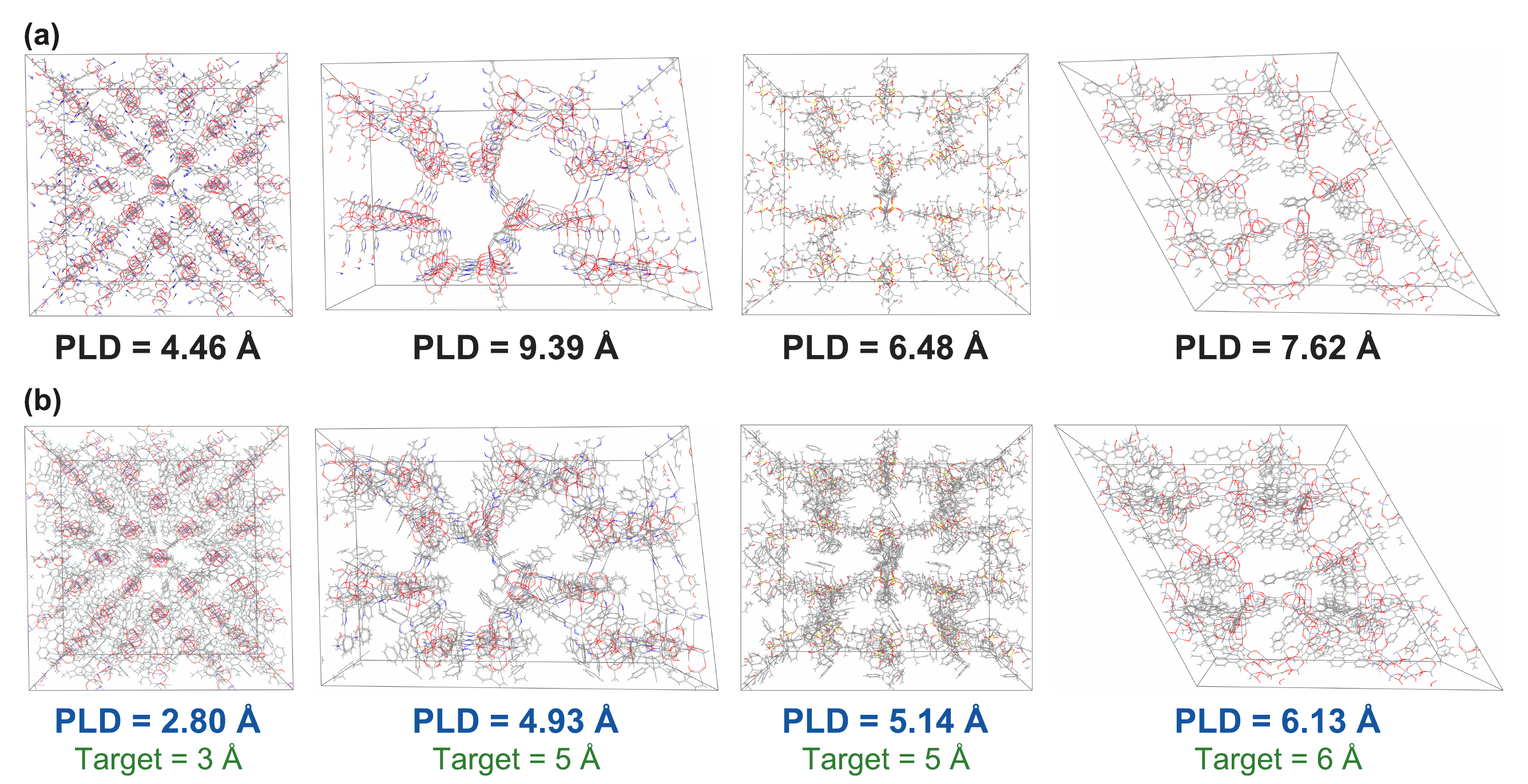}
\caption{\textbf{Editing-controlled pore-limiting diameter on representative MOFs.} (a) Source structures before editing; (b) the corresponding structures after CoFi-MOF editing, aligned column-wise. Under each structure, the measured PLD is reported in \AA.}
\label{fig:edit_cases}
\end{figure}

Figure~\ref{fig:edit_cases} illustrates the structural basis of local refinement: linker modifications alter the accessible aperture while preserving the surrounding topology and node identities. These examples complement the aggregate improvements above by showing how local chemical changes reshape the realized pore geometry.

\subsection{Downstream force-field screening of the candidate pool}
\label{app:relaxed_realizations}

As a downstream screening check, we relaxed candidates from the frozen CoFi-MOF pipeline and measured their apertures. The study contains 500 tasks, with 100 distinct initial noises at each target of 3--7~\AA. Model weights remain unchanged. Constructor candidates and MEF-edited candidates are assembled and measured before relaxation. For each task, candidates are ordered by assembled PLD error, with model rank and candidate identity breaking ties. We examine nested prefixes of this ordering, relax the corresponding structures with UFF4MOF, and measure relaxed PLD using Zeo++ \texttt{network -ha -res} throughout. The relaxation sequence comprises fixed-cell conjugate-gradient minimization, isotropic cell relaxation and FIRE minimization.

For each prefix, the reported output is the candidate with the smallest measured relaxed PLD error. All 500 tasks remain in the hit-rate denominator, and a task without a measurable relaxed structure counts as a miss. MAE uses tasks with a measurable relaxed output; its denominator therefore varies with the prefix. Confidence intervals use target-stratified task resampling with 5{,}000 bootstrap replicates. This analysis characterizes the candidate pool under the specified force-field and pore-measurement protocol.

\begin{table}[!htb]
\centering\small
\setlength{\tabcolsep}{5pt}
\caption{\textbf{Downstream force-field screening across model-generated realizations.} Each row examines a nested prefix of the assembled-geometry ordering and reports the closest measured relaxed PLD. Hit rates use all 500 tasks; MAE uses tasks with a measurable relaxed output.}
\label{tab:relaxed_realizations}
\begin{tabular}{rrrrr}
\toprule
Realizations examined & Relaxed Hit@0.5 (\%) & 95\% CI (\%) & Relaxed MAE (\AA) & Measurable tasks\\
\midrule
1 & 19.4 & 16.0--22.8 & 1.294 & 465\\
2 & 28.4 & 24.6--32.2 & 1.085 & 473\\
5 & 42.6 & 38.2--46.8 & 0.893 & 480\\
10 & 55.6 & 51.4--59.8 & 0.677 & 494\\
25 & \textbf{78.2} & \textbf{74.6--81.6} & \textbf{0.398} & 500\\
\bottomrule
\end{tabular}
\end{table}

Relaxed Hit@0.5 increases monotonically from 19.4\% to 78.2\%, while MAE decreases from 1.294 to 0.398~\AA. Examining 25 rather than one realization rescues 294 tasks and loses no existing hit, a paired gain of 58.8 percentage points with a 95\% confidence interval of 54.4--63.0 points. At 25 realizations, the numbers of hits at targets of 3--7~\AA{} are 89, 81, 77, 82 and 62 out of 100 tasks per target. The candidate pool therefore contains structures that meet the target under the tested force-field protocol, supporting downstream screening and experimental selection.

\subsection{Paired evaluation of learned linker editing}
\label{app:mef_control}
We evaluate MEF's proposal quality on 937 source-matched tasks at a target PLD of 6~\AA. Each task supplies the same editable source to MEF and random editing. Sources are selected from the verified Constructor pool, and all edited products are assembled and measured with Zeo++. Table~\ref{tab:mef_paired_control} also lists the direct output of the same attempt (w/o verifier and MEF), the globally highest-ranked assignment under the joint node--linker score without measured selection or editing.

\paragraph{First-proposal comparison.}
The first-proposal comparison evaluates the model's highest-ranked edit and the first randomly proposed edit without selecting by measured outcome. Assembly or measurement failures count as misses. MEF reaches 639/937 hits (68.20\%), compared with 315/937 (33.62\%) for random editing. The paired difference is 34.58 percentage points, with a 95\% bootstrap interval of [31.38, 37.89] from 5,000 paired resamples grouped by batch and source identity. Learned ranking doubles the first-proposal hit rate of random modification at the primary tolerance.

\paragraph{Verified best-of-ten comparison.}
With up to ten proposals per editor and the same measured-error selection over edited products, MEF reaches 78.12\% Hit@0.5 versus 65.21\% for random editing, with MAE of 0.3523 versus 0.5029~\AA. Learned ranking yields more target-accurate edits at an equal proposal budget.

\begin{table}[t]
\centering\small
\setlength{\tabcolsep}{4pt}
\caption{\textbf{Learned and random editing on 937 paired tasks at 6~\AA.} All tasks enter the hit-rate denominator, including failures. MAE and bias use measurable outputs and are reported in \AA. Both editors act on the same sources; w/o verifier and MEF shows the direct output of the same attempts.}
\label{tab:mef_paired_control}
\begin{tabular}{lrrrrrr}
\toprule
Output & Measured & Failed & Hits & Hit@0.5 (\%) & MAE & Bias\\
\midrule
w/o verifier and MEF & 906 & 31 & 140 & 14.94 & 1.8622 & $-0.6729$\\
\midrule
MEF first proposal & 932 & 5 & 639 & \textbf{68.20} & \textbf{0.4837} & $-0.3710$\\
Random first proposal & 923 & 14 & 315 & 33.62 & 1.0040 & $-0.7922$\\
\midrule
MEF best of ten & 937 & 0 & 732 & \textbf{78.12} & \textbf{0.3523} & $-0.2653$\\
Random best of ten & 937 & 0 & 611 & 65.21 & 0.5029 & $-0.3942$\\
\bottomrule
\end{tabular}
\end{table}

\paragraph{Distribution visualization.}
Figure~\ref{fig:mef_control}(a) uses the common subset of 890 tasks with measurable direct outputs, both first proposals and both best-of-ten outputs. Gaussian kernel densities use a shared 0.22~\AA\ bandwidth and reflection at zero. The 2--8~\AA\ display window retains all observations in density estimation without window-specific renormalization. In this subset, first-proposal hits are 613/890 for MEF and 309/890 for random editing, while the direct output reaches 136/890. The direct output has 20 observations below 2~\AA\ and 98 above 8~\AA; MEF and random first proposals have none below 2~\AA\ and one and eleven above 8~\AA, respectively. The plotted densities therefore describe a common measurable subset, whereas Table~\ref{tab:mef_paired_control} reports all-task performance.

\subsection{\revised{Learned ranking baselines}}
\label{app:ranking_baselines}

Linker editing is a learned ranker, so its quality can be attributed either to the model architecture or simply to learning on measured pore diameters. We separate the two with three simpler learned baselines trained on the same measured-PLD examples.

\paragraph{Task set and protocol.}
All rankers are evaluated on the same 1{,}129 tasks: the 500 tasks of the F500 attempt set and 629 F1000 tasks whose candidates carry cached measurements, of which 689 target 3~\AA{} and 110 target each of 4--7~\AA. Every method ranks the same candidate edits of the same editable source before any candidate PLD is read. Top-1 scores the highest-ranked edit without measurement. Top-10 applies the measured-error selection used elsewhere in the paper: up to ten proposals are verified with Zeo++ and the one closest to the target is kept. All 1{,}129 tasks enter the hit-rate denominator, and assembly or measurement failures count as misses. MAE is computed over tasks with a valid selected edit.

\paragraph{Baselines.}
The PLD-prediction MLP and the XGBoost regressor take the same source and edit features and regress the edited structure's measured PLD, ranking candidates by predicted distance to the target. The additive site-wise MLP sums independent site-level scores assigned to the proposed substitution sites. All three use the same measured-PLD training examples and average three seeds. MEF uses pretrained initialization, whereas the simpler learned baselines train from scratch; the comparison therefore covers these training configurations rather than isolating architecture under identical initialization. XGBoost uses depth 5, learning rate 0.05, subsampling 0.85 and early stopping on a held-out development split of 1{,}173 rows, which selects the number of boosting rounds.

\paragraph{Results.}
Table~\ref{tab:ranking_baselines} reports the comparison. MEF reaches 70.59\% Top-1 Hit@0.5, 31.21 points above XGBoost, the strongest of the three baselines on this metric, and 36.34 and 39.56 points above the two MLPs. Verification narrows the gap, since with up to ten measured proposals the baselines reach 73.31--76.23\%, but MEF retains the highest hit rate and the lowest target MAE: 0.2119~\AA{} against 0.2501~\AA{} for XGBoost and 0.2764 and 0.2342~\AA{} for the MLPs, a 15.3\% reduction relative to the strongest baseline. Measurement therefore recovers part of the difference in first-proposal quality, while learned ranking matters most when the measurement budget is small.

\begin{table}[t]
\centering\small
\setlength{\tabcolsep}{4pt}
\caption{\revised{\textbf{Comparison with learned ranking baselines on 1{,}129 matched tasks.} Top-1 scores the highest-ranked edit without measurement; Top-10 verifies up to ten proposals with Zeo++ and keeps the one closest to the target. Hits use all tasks as the denominator including failures; MAE uses tasks with a valid selected edit.}}
\label{tab:ranking_baselines}
\begin{tabular}{lrrrr}
\toprule
Ranker & Top-1 Hit@0.5 (\%) & Top-1 MAE & Top-10 Hit@0.5 (\%) & Top-10 MAE\\
\midrule
Full MEF & \textbf{70.59} & \textbf{0.3268} & \textbf{77.41} & \textbf{0.2119}\\
PLD-prediction MLP & 34.25 & 0.8240 & 73.31 & 0.2764\\
Additive site-wise MLP & 31.03 & 0.9293 & 76.23 & 0.2342\\
XGBoost PLD regressor & 39.39 & 0.6777 & 74.87 & 0.2501\\
\bottomrule
\end{tabular}
\end{table}

\subsection{\revised{Hit rate as a function of the evaluation budget}}
\label{app:budget_curve}

\begin{figure}[t]
\centering
\includegraphics[width=0.96\linewidth]{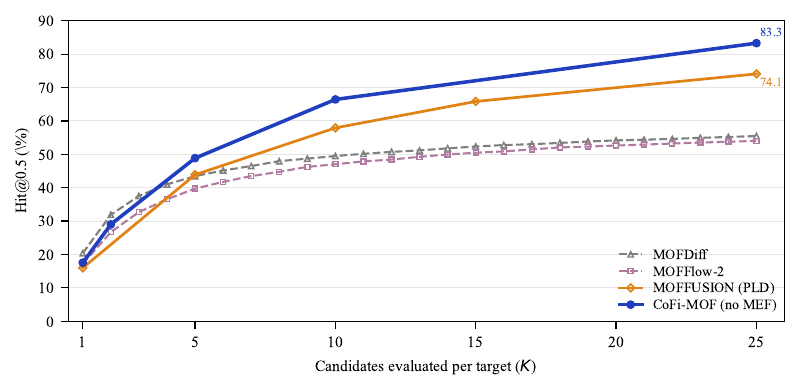}
\caption{\revised{\textbf{Hit@0.5 versus the evaluation budget $K$.} Each point averages the five targets of 3--7~\AA{} at equal weight. CoFi-MOF uses ranked Constructor candidates without MEF editing; all four methods are measured on the same target set under the shared 25-candidate protocol of Appendix~\ref{app:budget25}.}}
\label{fig:budget_curve}
\end{figure}

Figure~\ref{fig:budget_curve} reports Hit@0.5 as a function of the number of candidates evaluated per target for the four methods of Table~\ref{tab:main}. The CoFi-MOF curve is an offline replay of nested prefixes of the recorded ranked Constructor pool before any MEF editing over the same 5{,}000 allocated attempts, with failures counted as misses at targets of 3--7~\AA; the baseline curves are the matched best-of-$K$ re-evaluations described in Appendix~\ref{app:budget25}. $K$ is the nominal maximum number of candidate assignments considered, and failed assemblies reduce the number of Zeo++ calls, so the curves compare nominal budget rather than wall-clock cost.

Two features separate modeling quality from test-time search. First, every method improves steeply up to $K\approx5$ and then flattens, so a large part of each curve is search rather than generation: MOFDiff rises from 20.40\% to 43.48\% by $K=5$ and reaches only 55.48\% at $K=25$, and MOFFlow-2 behaves similarly (17.16\% to 39.72\% to 54.02\%). Second, CoFi-MOF's fixed-upstream realization curve exceeds MOFFUSION at $K=1$, overtakes the remaining baselines by $K=5$, and reaches 83.34\% at $K=25$ against 74.10\% for the strongest baseline. MOFFUSION also improves beyond $K=10$ (57.92\% to 74.10\%). These curves compare complete proposal-and-realization pipelines under the stated nominal budgets; they do not isolate the denoiser or equate candidate count with compute.

\subsection{\revised{Novelty relative to training and reference collections}}
\label{app:structural_novelty}

We compare 4,994 generated outputs against the full global corpus (all splits), a CoRE subset reconstructed in our assembly representation, and BW-DB at the recipe, linker-graph and crystal-structure levels (Table~\ref{tab:structural_novelty}).

\paragraph{Recipe and linker graphs.}
For 4,741 outputs (94.93\%), the topology--node--linker combination is absent from the global corpus. Among 1,698 selected MEF-edited outputs, 1,688 (99.41\%) have absent combinations. Comparing their edited linkers against 645 reference nonempty linker graphs finds no isomorphic match for 1,590 outputs (93.64\%). The graph comparison uses simplified heavy-atom connectivity and ignores bond orders, hydrogen atoms, charges and stereochemistry; the result describes novelty under this representation.

\paragraph{Crystal-structure comparison.}
We use \texttt{pymatgen StructureMatcher} with lattice tolerance 0.2, site tolerance 0.3 and angle tolerance $5^\circ$, enabling primitive-cell reduction and volume scaling. Table~\ref{tab:structural_novelty} reports the results. The CoRE reference contains 1,034 ASR, 637 FSR and 129 Ion records. There are 155 outputs with a composition represented in this CoRE subset and 601 with a composition represented in BW-DB; the two output groups may overlap. No crystal match is found under the stated protocol. Matcher checks recover all 20 self-matches and all 20 matches after fractional-coordinate perturbations of 0.02--0.03.

\begin{table}[t]
\centering\small
\setlength{\tabcolsep}{4pt}
\caption{\revised{\textbf{Novelty against specified reference collections.} Reference sizes count records. Same-composition counts refer to generated outputs with a composition present in the reference collection and can overlap across collections. A dash denotes a count not reported.}}
\label{tab:structural_novelty}
\begin{tabular}{lrrrr}
\toprule
Reference & Records & Outputs & Same composition & Matches\\
\midrule
Global corpus (all splits) & 136,911 & 4,994 & --- & 0\\
Reconstructed CoRE subset & 1,800 & 4,994 & 155 & 0\\
BW-DB & 308,962 & 4,994 & 601 & 0\\
\bottomrule
\end{tabular}
\end{table}

These no-match results apply to the specified references and settings. Crystal matching depends on representation, structural state and tolerances; the results do not establish novelty against all known MOFs.

\end{document}